\documentclass[
 reprint,=
 amsmath,amssymb,
 aps,unsortedaddress]{revtex4-1}
\usepackage{xcolor}
\usepackage{graphicx}
\usepackage{dcolumn}
\usepackage{bm}
\usepackage{amsmath}
\usepackage{subfigure}
\newcommand{\yy}{\it \color{red}}

\usepackage{ulem}

\begin{document}

\title{Emergent fractons and algebraic quantum liquid from plaquette melting transitions}

\author{Yizhi You}
\affiliation{Princeton Center for Theoretical Science, Princeton University, 
NJ, 08544, USA}

\author{Zhen Bi}
\affiliation{Department of Physics, Massachusetts Institute of Technology, Cambridge, MA 02139, USA}

\author{Michael Pretko}
\affiliation{Department of Physics and Center for Theory of Quantum Matter, University of Colorado, Boulder, CO 80309}

\date{\today}

\begin{abstract}
Paramagnetic spin systems with spontaneously broken spatial symmetries, such as valence bond solid (VBS) phases, can host topological defects carrying non-trivial quantum numbers, which enables the paradigm of deconfined quantum criticality.  In this work, we study the properties of topological defects in valence plaquette solid (VPS) phases on square and cubic lattices.  We show that the defects of the VPS order parameter, in addition to possessing non-trivial quantum numbers, have fracton mobility constraints deep in the VPS phase, which has been overlooked previously.  The spinon inside a single vortex cannot move freely in any direction, while a dipolar pair of vortices with spinon pairs can only move perpendicular to its dipole moment.  These mobility constraints, while they persist, can potentially inhibit the condensation of vortices and preclude a continuous transition from the VPS to the N\'eel antiferromagnet.  Instead, the VPS melting transition can be driven by proliferation of spinon dipoles. For example, we argue that a $2d$ VPS can melt into a stable gapless phase in the form of an algebraic bond liquid with algebraic correlations and long range entanglement. Such a bond liquid phase yields a concrete example of the elusive $2d$ Bose metal with symmetry fractionalization.  We also study $3d$ valence plaquette and valence cube ordered phase, and demonstrate that the topological defects therein also have fractonic dynamics. Possible nearby phases after melting the valence plaquettes or cubes are also discussed.
\end{abstract}

\maketitle
\tableofcontents

\section{Introduction}
The search for exotic quantum phases and their transitions has gained a wide audience in recent years, due to a range of unusual properties which cannot be understood within the Landau paradigm of symmetry breaking.  For example, topologically ordered phases, which do not possess a local order parameter, can exhibit topologically protected ground state degeneracies and deconfined fractionalized excitations.  Furthermore, even for seemingly conventional symmetry-breaking phases, it is possible to have Landau-forbidden phase transitions with similar unusual phenomenology.  The most well-known example of such a transition is between a N\'eel antiferromagnet and a valence bond solid (VBS).  These phases break two different types of symmetries (spin rotation and lattice symmetries, respectively), so Landau theory would predict that a generic transition between these phases is either first order, or possesses an intermediate regime where the two orders coexist.  In contrast, a more in-depth analysis reveals the existence of a generic second-order transition between these phases \cite{dqcp}.  While the N\'eel and VBS phases have fairly simple phenomenology, characterized by Landau order parameters, the unusual critical point between them hosts deconfined fractionalized excitations, earning it the name of a ``deconfined quantum critical point."  This phenomenon of deconfined quantum criticality has now been intensely studied in a variety of physical systems \cite{dqcp2,defined,levinsenthil,hedgehog,sandvik,melko,sandvik2,banerjee,sandvik3,harada,pujari,sandvik4,bartosch,banerjee2,nahumdeconfined,dualdeconfined,vbsdeconfined}.

Intuition for this unusual transition can be gained by studying the defects of each type of order, which carry quantum numbers associated with the other symmetry.  For example, a vortex of VBS order naturally carries spin-$1/2$, as we review in Section \ref{sec:review}.  Destruction of the VBS phase via condensation of these vortices then necessarily breaks spin-rotation symmetry, leading to an antiferromagnetic N\'eel phase.  This point of view leads to a general picture for deconfined quantum criticality, with implications well beyond the N\'eel-VBS transition.  While the VBS phase is the most commonly studied spatially ordered phase with spin quantum numbers associated with its topological defects, it is certainly not the only one.  Another order of this type, often overlooked in discussions of deconfined quantum criticality, is a valence plaquette solid (VPS), in which spins are entangled in clusters of fours, instead of the pairwise entanglement associated with a VBS phase \cite{rsvp,sheng,xu2008resonating,zhuwhite,hexdimer,zhaoplaq,albuq}. Both VPS and VBS order, as candidates of paramagnetic crystals, have been oberserved in a wide variety of materials \cite{zayed20174,zhao2018symmetry}.  Like a VBS defect, vortices of this plaquette order also carry spin-$1/2$, so we expect that destruction of the VPS phase via vortex proliferation will lead to a phase with broken spin-rotation symmetry, such as a N\'eel antiferromagnet.  However, as we demonstrate in detail, this simple picture of vortex proliferation is complicated by the fact that vortices of VPS order behave like fracton quasiparticles, with their characteristic mobility restrictions, leading to important consequences for phase transitions out of the VPS phase.

A fracton is a type of quasiparticle which does not have the ability to move by itself \cite{review}.  However, they can often move upon coming together to form certain bound states.  In the simplest cases, two fractons can move together upon forming a dipolar bound state.  Fractons were first encountered in the context of exactly solvable spin liquid models \cite{Vijay2015-jj,Vijay2016-dr,Chamon2005-fc,Haah2011-ny,yoshida2013exotic}, and have since been shown to have physical realizations ranging from crystalline defects \cite{elasticity,pai2018fractonic,gromov2017fractional,supersolid,kumarpotter} to hole-doped antiferromagnets \cite{polaron}.  Spin models for fracton systems have been intensely studied in recent years \cite{shirley2017fracton,paifusion,slagle2018foliated,shirley2018foliated,you2018symmetric,song2018twisted,Slagle2017-gk,Ma2017-qq,Ma2017-cb,Vijay2017-ey,Petrova2017-pe,Schmitz2017-ky,schmitz2018gauge,huang2018cage,rank2ice,compactify}, in large part due to their potential applications towards quantum information storage \cite{Haah2011-ny,bravyihaah,terhal}.  Fracton models have also exhibited unexpected connections with various other areas of physics, ranging from gravitation \cite{gravity,holo} to many-body localization \cite{Chamon2005-fc,Prem2017-ql,frc}, and have even manifested in the theory of deconfined quantum critical points between certain VBS phases \cite{ma2018higher}.  Theoretically, fractons are often described in the language of symmetric tensor gauge theories, which encode the immobility of fractons in a set of higher moment conservation laws, such as conservation of dipole moment \cite{pretko2017subdimensional,pretko2017generalized,ma2018fracton,bulmash2018higgs,bulmash2018generalized,Pretko2017-nt,fractoncs,gromov2018towards}.

The tensor gauge theory description allows us to make an immediate connection between plaquette order and the physics of fractons.  As we review in Section \ref{sec:review}, a conventional valence bond solid can be mapped onto the confined phase of a vector gauge theory, with vortices behaving as linearly confined charges.  In similar fashion, we show in Section \ref{sec:2d} that a VPS phase can be mapped onto the confined phase of a symmetric tensor gauge theory, with vortices of the plaquette order acting as the fractons.  Even though these fractons are confined within the VPS phase (corresponding to the large energy cost associated with vortices), their mobility restrictions still can have important consequences for phase transitions driven by vortex proliferation.  For a valence bond solid, the spin-carrying vortices become gapless deconfined excitations at a quantum critical point, then subsequently condense to drive the system into a N\'eel antiferromagnet.  For a VPS phase, on the other hand, the vortices are immobile fractons, for as long as the description in terms of plaquettes remains valid.  (The mobility restrictions could break down in a regime where a plaquette can easily break down into a pair of dimers.)  Assuming we remain in a regime well-described in terms of plaquettes, potential phase transitions and quantum critical points can be strongly impacted by the fractonic nature of the vortices.  Even if the fractons become deconfined at a quantum critical point, their mobility restriction serves as an impediment to direct condensation.  Fracton systems therefore have a tendency to first exhibit a condensation transition of mobile dipoles, which relaxes the mobility restrictions and allows a subsequent fracton condensation transition.  This two-stage nature of fracton condensation transitions is dramatically realized in the analysis of two-dimensional quantum melting, which predicts that two-dimensional crystals must pass through a hexatic phase before fully melting to an isotropic liquid \cite{elasticity,supersolid,kumarpotter}.

We therefore conclude that VPS order can generically host an intermediate phase in which only \textit{dipoles} of vortices are condensed, while individual vortices remain gapped.  The precise nature of this intermediate phase depends on the microscopic details governing the interaction of two vortices within a dipole pair.  In certain cases, the intermediate phase may be a simple bond-ordered phase, such as a valence bond solid.  As another illustrative example, we argue that a transition between two-dimensional VPS and N\'eel phases can feature a stable intermediate gapless phase in the form of an algebraic bond liquid \cite{xu2007bond,paramekanti2002ring,tay2010possible,tay2011possible} with quasi-long-range order between dipoles formed by spinon pairs. Such a featureless gapless liquid carries many features akin to the concept of the `Bose metal,' including power law correlations and zero-energy nodal lines. In particular, its thermodynamic and entanglement properties have the hallmarks of a $2d$ Fermi liquid and hence can be regarded as the `boson descendant' of a Fermi surface\cite{mishmash2011bose}.
A gapless intermediate phase of this type is consistent with existing numerics on the VPS-N\'eel transition in 2d Heisenberg models \cite{sheng}.  In Section \ref{sec:2d}, we review the stability of this algebraic bond liquid phase and study its transition with a VPS phase. We also describe some novel properties which may be used to detect an algebraic bond liquid in experiments or numerics, such as specific heat, structure factor, and entanglement entropy. The bond liquid exhibits novel characteristics, including $T\ln(T)$ dependence for specific heat\cite{paramekanti2002ring} and long-range entanglement, with entanglement entropy scaling as $L\ln(L)$ which exceeds the boundary law\cite{lai2013violation}.

In Section \ref{sec:3d}, we extend these ideas to three-dimensional cube-ordered and valence plaquette phases on a cubic lattice.  In each case, the fundamental topological defects of the order behave as immobile fractons.  In the case of cube order, even spinon dipoles are locked in place, with spinon quadrupoles behaving as one-dimensional lineons.  We also study different types of plaquette order in three dimensions, focusing on a strongly anisotropic VPS phase which can exhibit a continuous transition to a VBS phase.  This phase transition features a reduced dimension allowing the physics of certain 3-dimensional VPS melting transitions to be mapped onto the problem of a 2-dimensional VBS melting transition.

\section{Review of the VBS-N\'eel transition}
\label{sec:review}

Let us first recall the prominent deconfined quantum critical point between VBS and N\'eel phases on a 2D square lattice spin-$\frac{1}{2}$ system. For an interacting spin-$\frac{1}{2}$ model on a square lattice, the Lieb-Schultz-Mattis theorem forbids any featureless gapped paramagnet ground state \cite{lsm,hastings,oshikawa}. Two common symmetry breaking states are the N\'eel order, which spontaneously breaks spin rotation and lattice translation symmetries, and the valence bond solid (VBS) order, which breaks the lattice translation and $C_4$ rotation. 

\begin{figure}[h]
  \centering
      \includegraphics[width=0.4\textwidth]{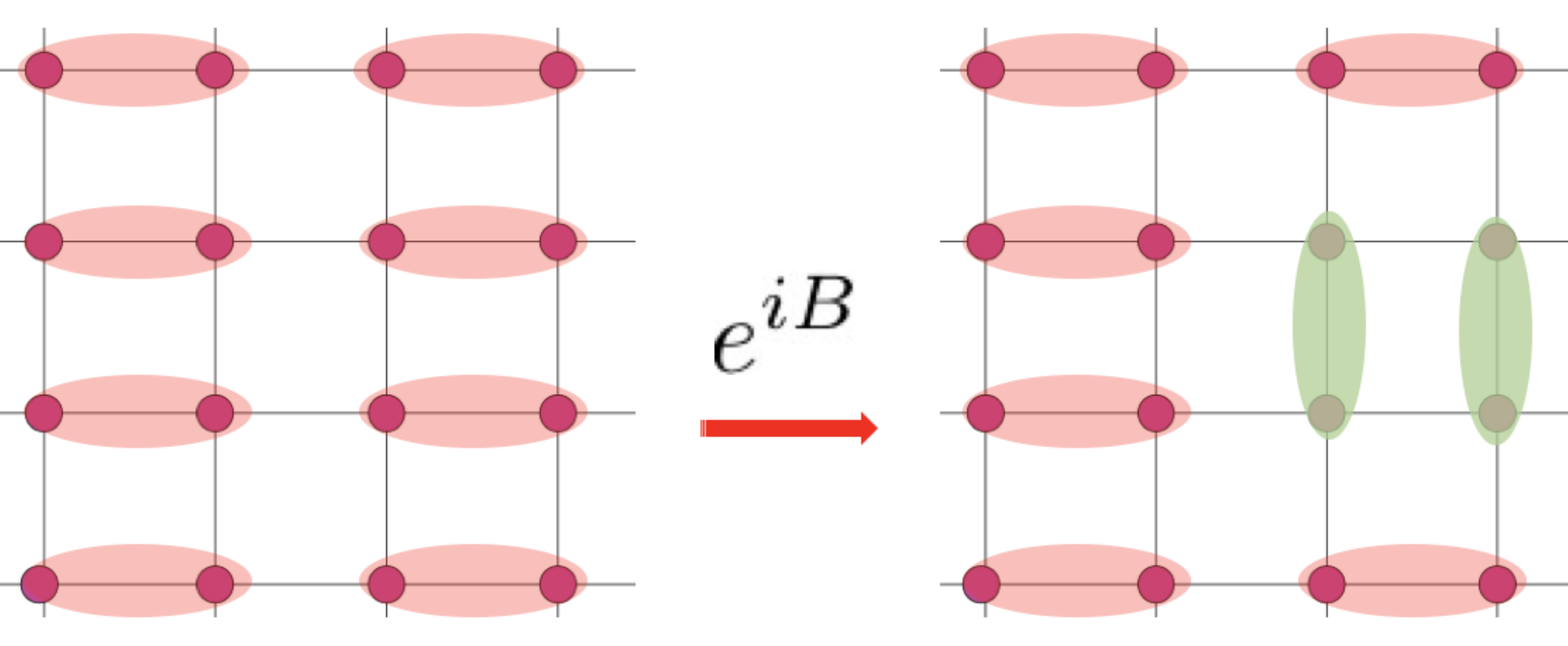}
  \caption{A typical VBS order on a square lattice. When mapping the dimer order to a compact U(1) gauge theory, the magnetic flux operator flips local VBS configurations.
}
\label{vbs1}
\end{figure}
 
\begin{figure}[h]
  \centering
      \includegraphics[width=0.47\textwidth]{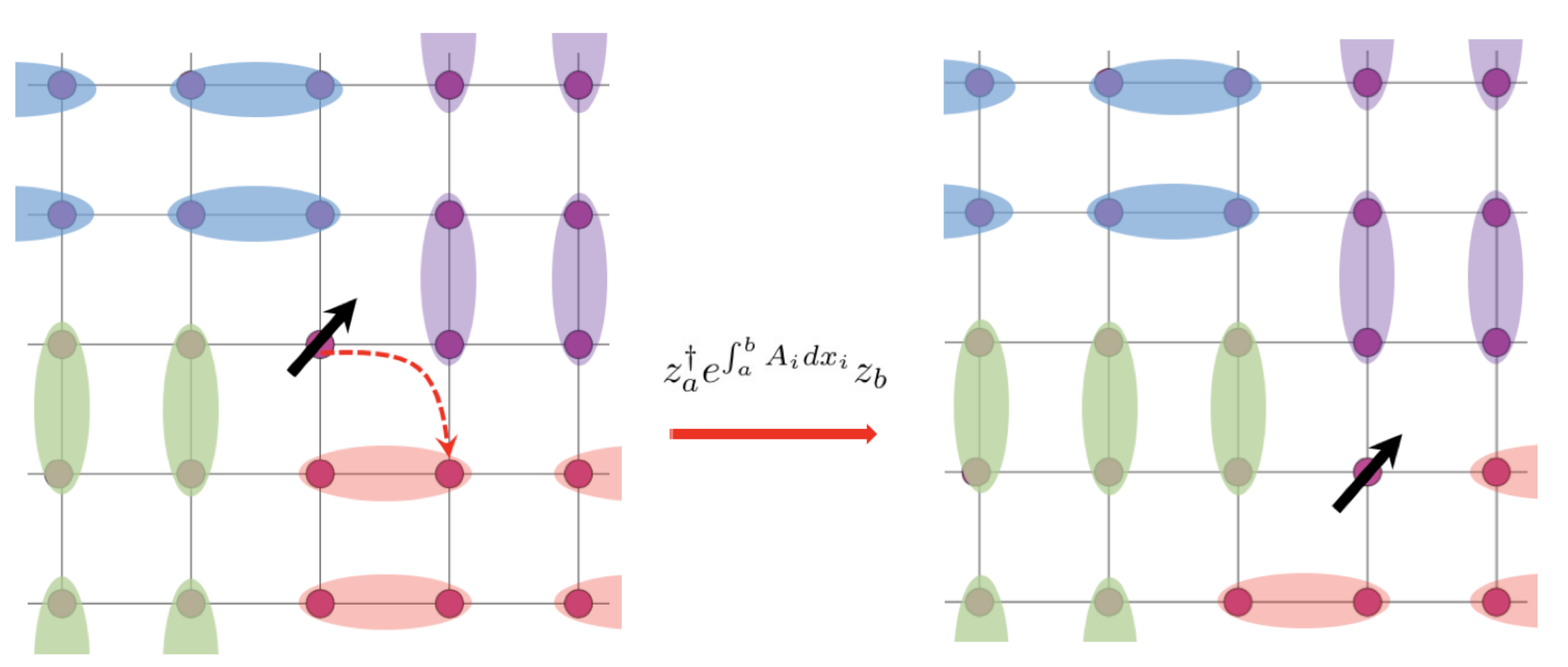}
  \caption{VBS vortex with a spinon inside the vortex core. The spinon can fluctuate locally by VBS pattern reconstruction. This implies the spinon current carries a gauge charge with respect to the VBS order.
} 
\label{vbs2}
\end{figure}

 In Ref.~\cite{dqcp,dqcp2,defined}, it was proposed that, despite their distinct symmetry breaking patterns, the VBS and N\'eel phases in a square lattice spin-$\frac{1}{2}$ system can be connected by a continuous phase transition which is dubbed as a deconfined quantum critical point. The key ingredient for such a Landau-forbidden phase transition between two symmetry broken phases lies in the fact that the topological defect of the order parameter for one broken symmetry carries a nontrivial quantum number of the other symmetry group.  In order to restore a broken symmetry, one can consider the condensation of topological defects of the order parameter. However, this condensate would drive the system into another broken symmetry state simultaneously due to the charge carried by the defect. In the VBS-N\'eel example, the VBS phase spontaneously breaks the $C_4$ lattice rotations, which has four degenerate ground state patterns related by the $C_4$ rotations, as depicted in Fig.\ref{vbs2}. The VBS order can be characterized as a complex scalar with four-fold anisotropy. To quantum disorder the VBS order, one should proliferate the $Z_4$ vortices. Since the VBS vortex carries an unpaired spin, as shown pictorially in Fig. \ref{vbs2}, consequently the vortex condensation breaks spin rotation symmetry. 

To formulate the VBS-N\'eel transition, it is useful to note that the VBS phase can be mapped to the confined phase of a compact $U(1)$ gauge theory. The dimer coverage on the edge of the square lattice can be mapped to an electric field,
\begin{align} 
E_i(\bm{r})=(-1)^{i_{\bm{r}}} D_i(\bm{r})
\end{align}
where $D_i(\bm{r})$ refers to the dimer coverage on the edge adjacent to site $\bm{r}$ along $i$-direction, which only takes values $0$ or $1$. The index $i_{\bm{r}}=0$ or $1$ for $\bm{r}\in$ A or B sublattice respectively. In the VBS phase, each spin can be paired with one and only one of its neighboring spins to form spin singlets. This translates to the local Gauss's law for the electric fields,
\begin{align} 
\partial_i E_{i}(\bm{r})=(-1)^{i_{\bm{r}}} (1-q(\bm{r}))
\label{eq:Gauss}
\end{align}
where $q(\bm{x})$ is the number of unpaired spinons at site $\bm{r}$. (It is worth noting that the `spinon quantum number' corresponds to the gauge charge of the emergent U(1) gauge field carried by the fractionalized spinon, which should be distinguished from the $S_z$ charge.)  A free spinon charge appears if no dimer is touching the site. Breaking a dimer can create a pair of spinons.  Further, let us define the gauge connection, $A_i$, which is the conjugate variable to $E_i$, namely $[A_i(\bm{x}),E_j(\bm{y})]=\frac{i}{2\pi}\delta_{ij}\delta_{\bm{xy}}$. In the spin picture, $A_i$ operator breaks (creates) a dimer on link-$i$ if the link does (not) have an existing dimer. The magnetic flux operator $B=\nabla \times A$ flips local dimer orientation on a plaquette, as in Fig.~\ref{vbs1}. Thus, the low-energy physics of the VBS phase is characterized by a compact $U(1)$ gauge theory with background charges whose Hamiltonian is given by the following,
\begin{align} 
\mathcal{H}&=U\sum_{\bm{r},i} E_i(E_i-(-1)^{i_{\bm{r}}})+ T\sum_{\square}\cos(\nabla\times A)\\
&+V\sum_{\bm{r}}(\partial_iE_i-(-1)^{i_{\bm{r}}})^2
\end{align}
The pure (2+1)d compact $U(1)$ gauge theory, due to the instanton events, is always in its confined phase at low energy, where the spinon excitations experience a linearly confining potential. The confined phase is mapped to the VBS phase where the VBS vortices have linear confinement due to the energy cost of the domain walls connecting the spinons.

Consider the phase transition out of the VBS state by proliferating the VBS vortices. The quantum critical point can be described by the following field theory,
\begin{align} 
\mathcal{L}=|(\partial_{\mu}-iA_{\mu})z|^2+r|z|^2+g|z|^4+\frac{1}{4e^2}F^2+...,
\label{eq:DQCP}
\end{align}
where $z$ is a $CP^1$ field that captures the spinon degree of freedom inside the vortex core. $A_{\mu}$ is the emergent gauge field described previously.  Since the spinon is charged under the emergent gauge field as in Eq. \ref{eq:Gauss}, they are minimally coupled to the gauge fields. The hopping of a spinon thus requires the change of dimer configurations along the path.
In this theory, we implicitly include the 4-fold monopole creation and annihilation operators which correspond to the possible instanton events that respect the lattice $C_4$ rotation symmetry\cite{dqcp,dqcp2,defined}. When the spinon is gapped, namely $r>0$, the gauge theory is confined due to monopole proliferation and the system is in the VBS phase. At the critical point, $r=0$, the spinon field becomes massless and there is evidence that the compact $U(1)$ gauge field dynamically becomes deconfined at the fixed point\cite{dqcp}, $i.e.$ the 4-fold instanton events are irrelevant under renormalization group flow. As $r$ decreases below 0, the VBS vortices/spinons condense which restores the $C_4$ rotation and spontaneously breaks the spin rotation symmetry.

\section{Plaquette paramagnet in 2D}
\label{sec:2d}

\begin{figure}[h]
  \centering
      \includegraphics[width=0.43\textwidth]{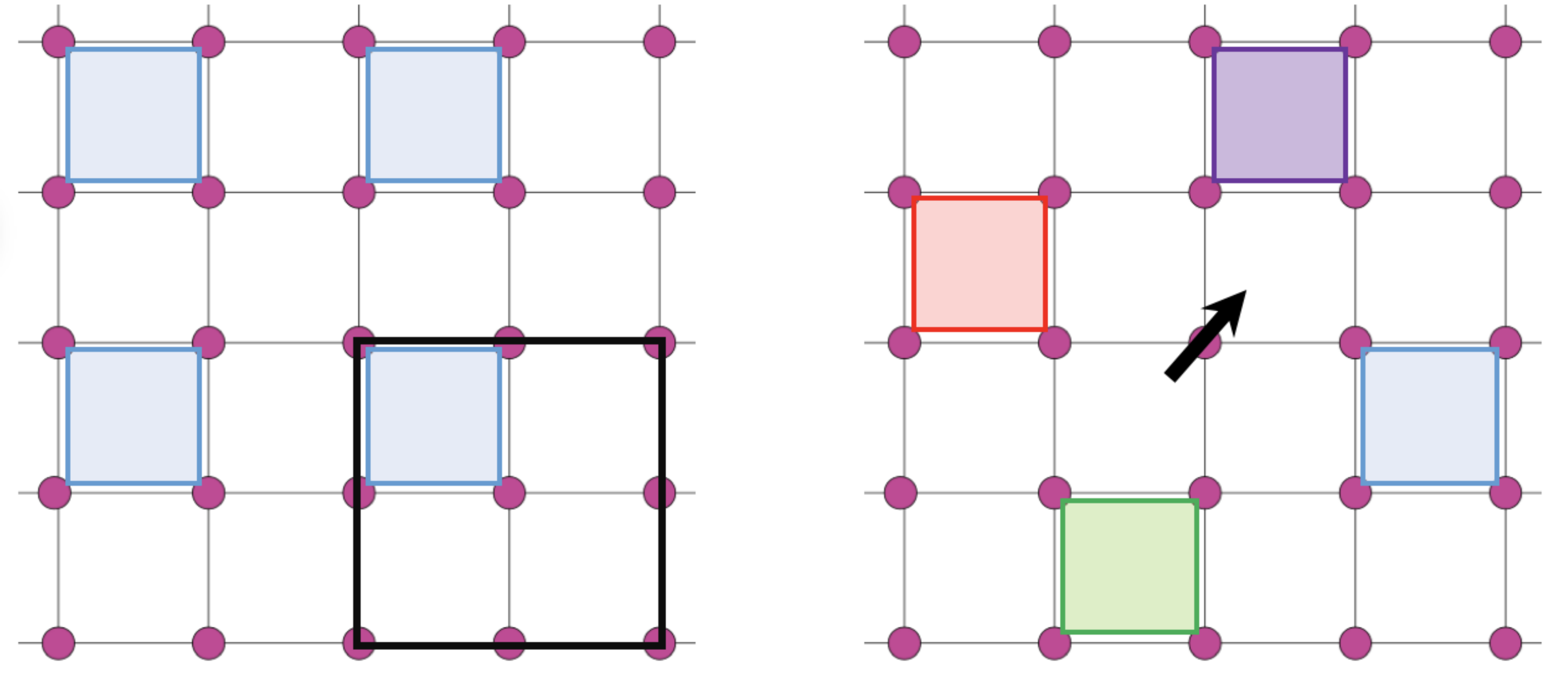}
  \caption{Left: VPS order which enlarges the unit cell by 4. Right: The vortex connecting four distinct VPS patterns carries a spinon.} 
\label{plaq1}
\end{figure}

Apart from columnar valence bond order, another widely observed paramagnetic crystalline phase is the VPS (valence plaquette solid) state which breaks $C_4$ symmetry and lattice translation $T_x,T_y$ for both directions. Such a plaquette paramagnet has been found and fabricated in frustrated magnets and AMO systems\cite{zayed20174,xu2008resonating,pankov2007resonating}.  There are various types of VPS wave functions which respect the same symmetries.  For example, for a spin-$1/2$ system, each valence plaquette represents a symmetric combination of vertical and horizontal dimer pairs on the same plaquette. For an $SU(4)$ spin system with a fundamental representation on each site, one can have plaquettes in an entangled state of four $SU(4)$ spins.  Regardless of the microscopic configuration inside the valence plaquette, each spin participates in only one of the four plaquette clusters adjacent to the site. The VPS order enlarges the unit cell into four plaquettes, so there are four distinct VPS patterns related by site-centered $C_4$ rotation, as shown in Fig.~\ref{plaq1}.

\subsection{Defect structure}

\begin{figure}[h]
  \centering
      \includegraphics[width=0.43\textwidth]{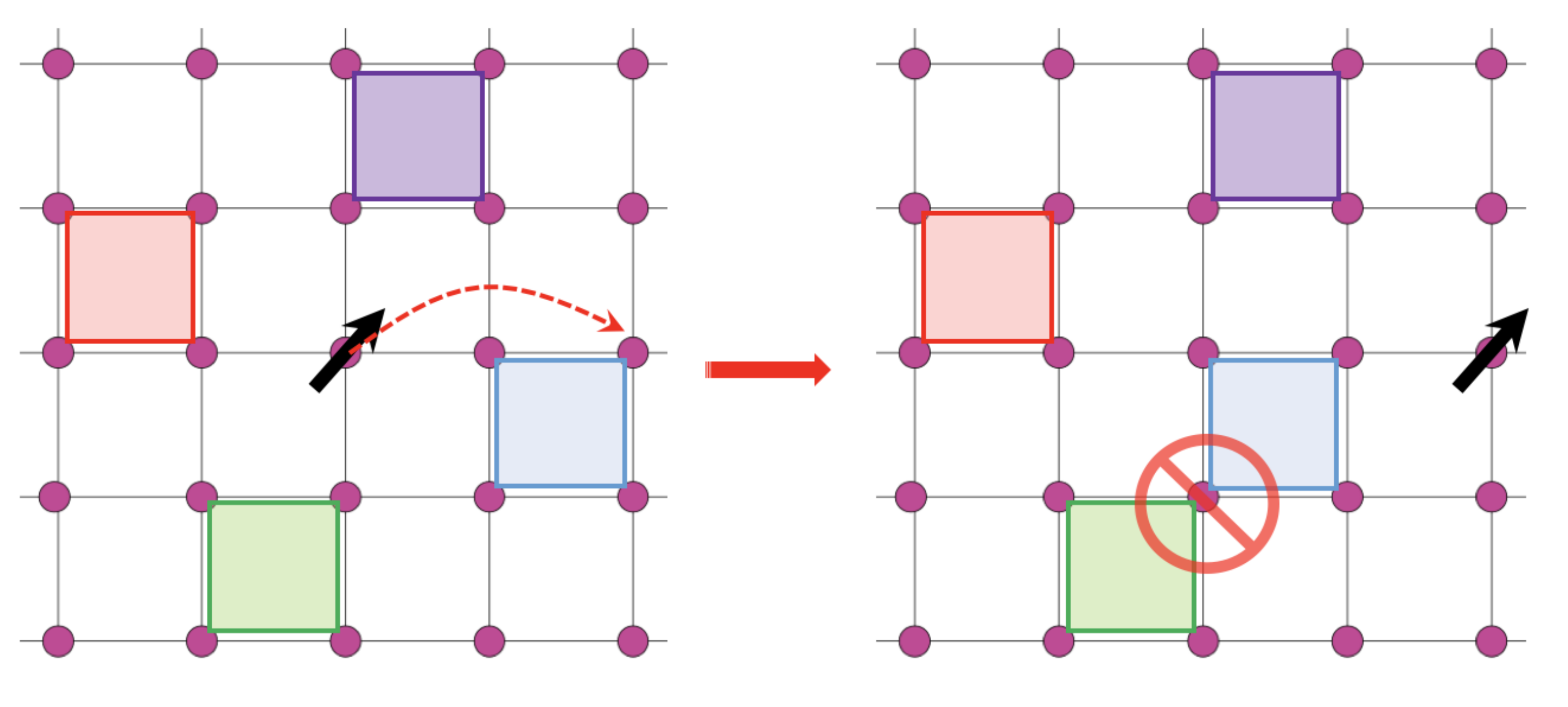}
  \caption{The spinon inside the VPS vortex has restricted mobility. It cannot move without breaking additional plaquettes.
} 
\label{plaq2}
\end{figure}

\begin{figure}[h]
  \centering
      \includegraphics[width=0.45\textwidth]{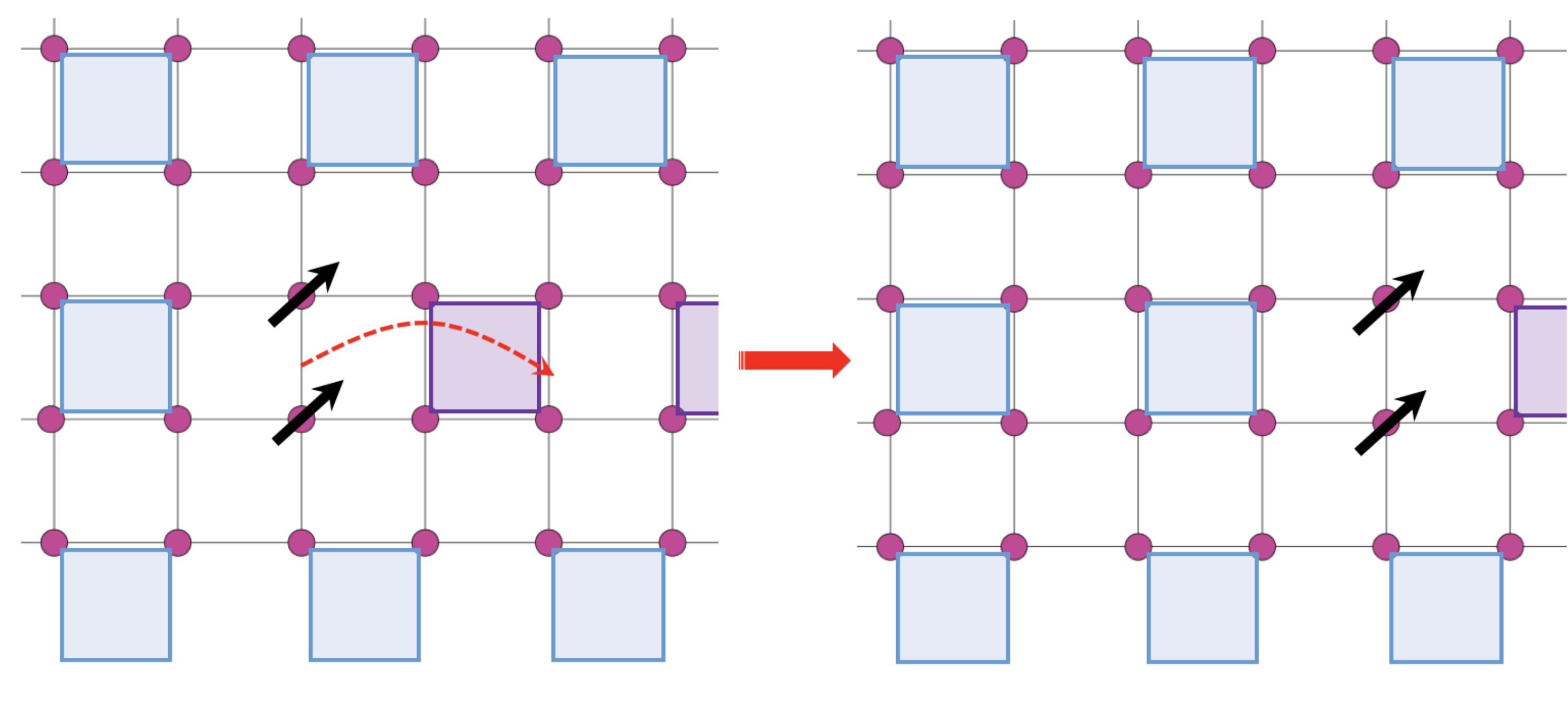}
  \caption{ A dipole can move along the stripe transverse to the dipole's orientation by exchanging position with a plaquette.
} 
\label{plaq3}
\end{figure}

To restore the spatial symmetry, it is essential to proliferate the $C_4$ symmetry defect. For the plaquette crystalline phase, one can define the four distinct plaquette patterns as a $Z_4$ boson. The $C_4$ symmetry defect formed by the vortex configuration of the order parameter carries a spinon. Naively, one expects the vortex condensate will restore the crystalline symmetry and concurrently develops magnetic order. However, in order to drive the defect condensate, vortices need to be able to fluctuate in spacetime.  As opposed to the VBS phase, where a spinon in the background of dimers can hop among sites by reconstructing local valence bond configuration, a spinon in the background of plaquette order is frozen - cannot move away from the original vortex center without breaking additional plaquettes, as depicted in Fig.~\ref{plaq2}.
 In contrast, a pair of spinons living on the link between adjacent sites can hop along the stripe perpendicular to that link without breaking additional plaquettes, as depicted in Fig.~\ref{plaq3}. Such a spinon pair, which we refer to as a spinon dipole, is a $1d$ subdimensional particle which moves transversely to the dipole's orientation.

Based on these observations, the topological defect of the plaquette order displays restricted motion which exactly resembles the behavior of fractons. In fracton phases of matter, the fundamental deconfined quasiparticle excitation is immobile due to a finite energy barrier associated with the creation of additional excitations.  Meanwhile, a pair of fractons (dipole) has a certain degree of mobility, though it too is often restricted to motion only within a submanifold, such as a line, plane, or fractal.

To make the connection between VPS defects and fractons precise, we introduce a higher rank gauge theory description for the valence plaquette order on a square lattice. By analogy to the VBS order, the plaquette order can be mapped to a rank-2 symmetric tensor electric field defined at the center of each square as the following. 
\begin{align} 
E_{xy}(\bm{r})=(-1)^{i_{\bm{r}}} P(\bm{r})
\end{align}
where $P=1\,(0)$ corresponds to the valence plaquette occupancy (vacancy) on each square. The index $i_{\bm{r}}$ is the same as defined before. As opposed to the VBS state, where dimers can have two orientations corresponding to $E_x$ and $E_y$, the plaquette electric field is a single-component field, effectively a scalar. We can also define a conjugate variable $A_{xy}$, satisfying $[A_{xy}(\bm{r}),E_{xy}(\bm{r'})]=\frac{i}{2\pi}\delta_{\bm{r},\bm{r'}}$.  The operator $e^{\pm iA_{xy}}$ creates/annihilates a valence plaquette.  As each spin on the site is only entangled with one of the four adjacent plaquette clusters, one can define a Gauss's law for the rank-2 electric field as,
\begin{align} 
\partial_x \partial_y E_{xy}(\bm{r})=(-1)^{i_{\bm{r}}} (1-q(\bm{r}))
\label{gauss}
\end{align}
where $q(\bm{r})$ is the number of unpaired spinon at site $\bm{r}$.  As long as there is one plaquette adjacent to a site, there is no free spinon on that site.  If plaquettes are absent from all four squares surrounding the site, then there exists a free spinon charge at the center.  This Gauss's law is precisely the two-dimensional version of the Gauss's law seen in the fracton phase of matter described by a hollow rank-2 symmetric tensor gauge theory \cite{ma2018fracton,you2018symmetric,bulmash2018higgs}.  Due to the particular double derivative in Eq.~\ref{gauss}, the spinon number is conserved on each row and column of the system, so the theory respects an emergent subsystem $U(1)$ symmetry:
\begin{equation}
\int dx\,q = 1-(-1)^y\int dx\, (-1)^x\partial_x\partial_y E_{xy} = \textrm{const.}
\end{equation}
A similar equation holds in the $y$-direction.  Again, we emphasize that $q$ is the spinon charge quantum number corresponding to the emergent $U(1)$ symmetry, which should be distinguished from the total $S_z$ charge.  In particular, only the spinon charge is conserved on row/columns while the $S_z$ charge is conserved globally.  Due to the emergent subsystem symmetry, single spinon motion is prohibited. However, a pair of spinons, which we refer to as a dipole, can hop only along the stripe perpendicular to its orientation.

\begin{figure}[h]
  \centering
      \includegraphics[width=0.25\textwidth]{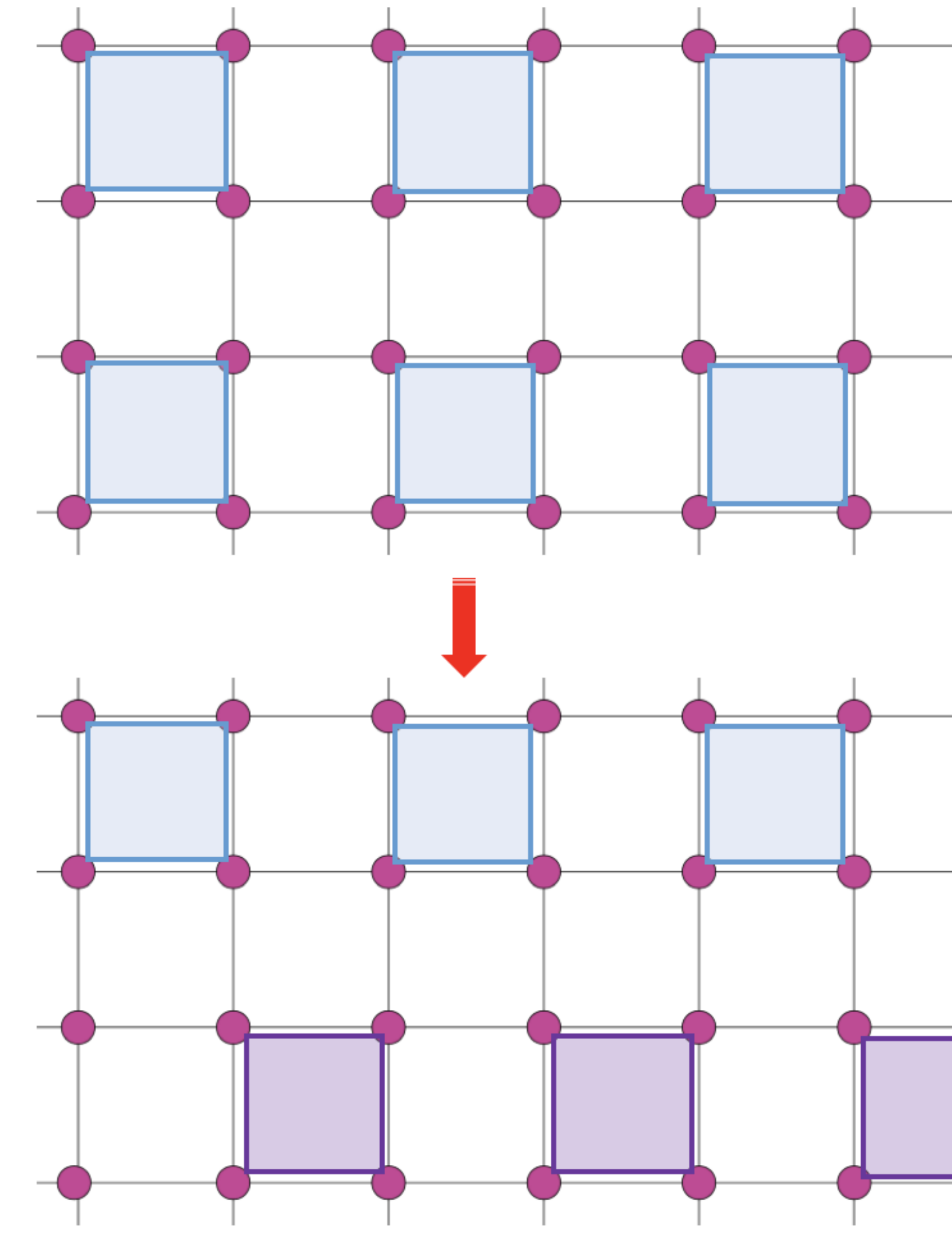}
  \caption{A global flux $\phi^x(y_i)$ globally shifts the valence plaquette configuration on the row with $y=y_i$.
} 
\label{plaq4}
\end{figure}

Based on the Gauss's law in Eq.~\ref{gauss}, the low-energy sector is invariant under the following gauge transformation,
\begin{align} 
A_{xy}  \rightarrow A_{xy}+\partial_x\partial_y \alpha,
\end{align}
for any function $\alpha$ with arbitrary spatial dependence.  Since there is only one component of the gauge field, there is no local gauge invariant operator representing a flux degree of freedom.  The absence of a local flux operator indicates that there is no resonant process that can flip plaquette configurations locally into each other.  However, one can define global flux operators,
\begin{align} 
\phi^x(y_i)=\int dx  (-1)^x A_{xy}(y_i),\\
\phi^y(x_i)=\int dy  (-1)^y A_{xy}(x_i) 
\label{eq:Gflux}
\end{align}
These global flux operators shift a row (or column) of the valence plaquette configuration by one unit cell, as in Fig.~\ref{plaq4}.

\subsection{Bond ordered phases}

Due to the restricted mobility of the spinon corresponding to the VPS defect, it is difficult for the spinon to condense directly.  We therefore do not expect a continuous transition from the VPS phase to a simple N\'eel antiferromagnet.  Based on this observation, our next goal is to figure out the possible phases near the plaquette melting transition, driven by condensation of dipolar pairs of vortices.  Since these vortex pairs live along the links of the lattice, it is natural to expect that the resulting phase will be well-described in terms of valence bonds, instead of valence plaquettes.  The precise nature of this intermediate phase will depend on the microscopic structure of the dipolar pair, which is dictated by the details of the underlying Hamiltonian.

In the simplest case, the spinon dipole may form an $SU(2)$ singlet.  In this situation, the result of dipole condensation will be a type of valence bond solid.  Even in this case, there are several types of possible VBS states which can result, depending on how precisely the dipoles condense.  For example, condensation of $y$-directed dipoles will naturally lead to a VBS state with all valence bonds aligned with the $y$-direction.  This physics is borne out by studies on two-dimensional quantum dimer models on a square lattice, which can host a continuous transition between plaquette order and a staggered VBS phase, whose critical point is described by the quantum Lifshitz theory with $z=2$ dynamical exponent \cite{cantor}.

\begin{figure}[h]
  \centering
      \includegraphics[width=0.25\textwidth]{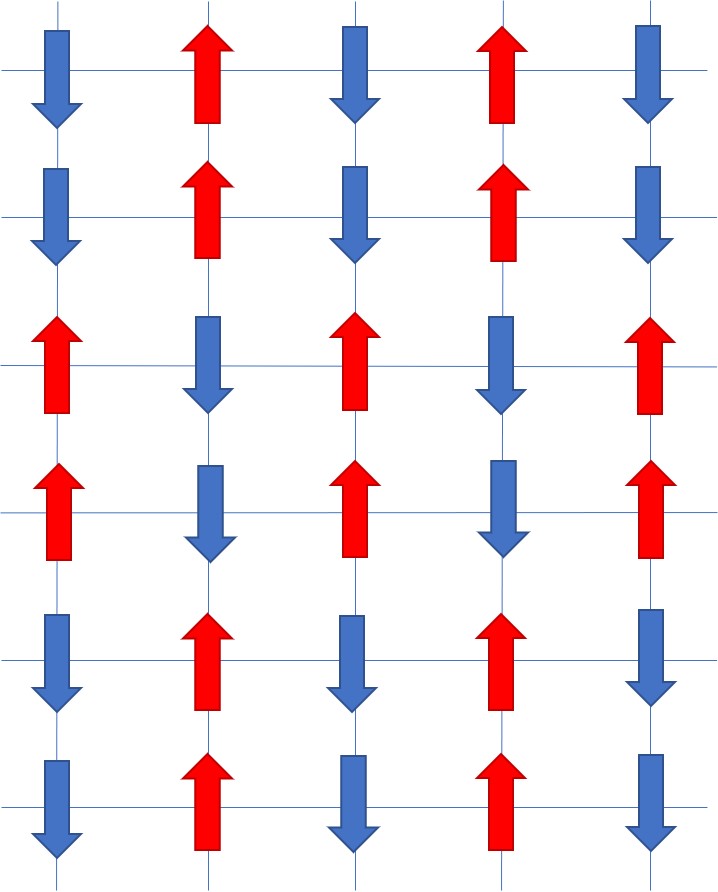}
  \caption{Condensation of triplet dipoles can lead to antiferromagnetic order coexisting with a form of bond order.
} 
\label{bondneel}
\end{figure}

Beyond such VBS phases, obtained by condensing singlet spinon dipole, another possibility is that the two spinons of a dipole tend to align their spins in a triplet state.  In this case, since the dipoles are carrying a net spin, the resulting condensed phase breaks spin rotation symmetry.  However, this state will not be the simple N\'eel phase, but rather will have antiferromagnetic order coexisting with a form of bond order.  An example of a system with this type of order is depicted in Figure \ref{bondneel}.  Further condensation of defects of the bond order can then drive this phase into a simple N\'eel antiferromagnet. 

\subsection{Algebraic bond liquid phase}

Beyond analytical exploration of transitions to bond-ordered phases, there are also various numerical simulations of frustrated spin models and hardcore boson models which appear to reveal a possible symmetric gapless phase nearby the VPS phase.  This indicates that a plaquette ordered system may melt into a more exotic phase under certain conditions.  In this section, we propose a stable algebraic bond liquid phase with power-law correlations which can potentially emerges near the VPS phase.

In our previous discussion, we have elucidated that the spinon pair along a link can hop and fluctuate along the $1d$ stripe perpendicular to the dipole orientation. During the melting of the VPS paramagnet, the dipole proliferates and fluctuates along the stripe with quasi-one-dimensional dispersion.  Because of the restricted dimension, the spinon dipole cannot establish long range order.  Instead, the dipoles form a state with quasi-long range order. The system form an algebraic bond liquid.

To elucidate the behavior of the emergent spinons during the plaquette melting transition, we implement the Schwinger boson representation,
\begin{align} 
&\bm{S}=\frac{1}{2}z^\dagger \bm{\sigma} z,
\label{eq:cp1}
\end{align}
with the constraint $z_1^\dagger z_1+z_2^\dagger z_2=1$. The two component field $(z_1,z_2)^{T}$ are the bosonic spinons. The total charge density is fixed in the original microscopic model. The relative density $\frac{1}{2}(z_1^\dagger z_1-z_2^\dagger z_2)$ corresponds to the $S_z$ quantum number.  Both $z_1$ and $z_2$ couple to the same emergent $U(1)$ gauge field, while each of them also carries $\pm 1/2$ charge for $S_z$. The magnon excitation $S^{+}=z_1^{\dagger} z_2$ corresponds to the exciton pair of Schwinger bosons with opposite flavors.  The subsystem charge conservation law in Eq.~\ref{gauss} indicates the gauge charge of the Schwinger boson is conserved on each line.  We can approximately express the Schwinger boson operator in terms of a $U(1)$ rotor field as $z^{\dagger}_a\sim e^{i \theta_a}$, writing the effective Hamiltonian of the bosons in the following form,
\begin{align} 
\mathcal{H}=&U\sum_{\bm{r}}E_{xy}(E_{xy}-(-1)^{i_{\bm{r}}})\\ \nonumber
&+u\sum_{\bm{r}}(\sum_{a=1,2}\hat{n}_a-1)^2+t \sum_{a=1,2}\cos(\partial_x \partial_y \theta_a+A_{xy}),
\label{eq1}
\end{align}
where the $u$ term is to implement the onsite constraint $n_1+n_2=1$, which can be released when coarse graining in the continuum limit.

Note that the Hamiltonian contains no usual kinetic term for the phase variables such as $(\partial_i \theta_a)^2$, as a single spinon is immobile in any direction. Instead, the leading order kinetic term $\partial_x \partial_y \theta$ corresponds to an $x$-directed dipole hopping along the $y$ direction, or vice versa. Such dipole hopping requires reconstruction of the valence plaquette pattern along the hopping path, so the dipole current minimally couples with the gauge field $A_{xy}$. Since there is no local gauge flux for such a higher rank gauge field, the role of gauge fluctuation merely projects the Schwinger boson to the physical Hilbert space of $n_1+n_2=1$. 


In the easy-plane anisotropy limit where $\langle S_z \rangle$=0, we take $\langle n_1 \rangle=\langle n_2 \rangle=\frac{1}{2}$ so each slave boson is at half-filling.
This theory resembles the exciton bose liquid phase studied in Ref.~\cite{xu2007bond,paramekanti2002ring,tay2010possible,tay2011possible}, where the bosons interact via a ring-exchange interaction which respects a subsystem $U(1)$ symmetry. Such an exciton bose liquid constitutes a stable gapless phase with power-law correlation at fractional filling.  When the $u/t$ becomes small and the theory tends to have order in $\theta$ variables, we can implement a ``spin-wave" approximation, expanding the cosine terms, and obtain a Guassian theory,

\begin{align} 
\mathcal{L}=\frac{K}{2} \sum_{a=1,2} (\partial_t \theta_a)^2-\frac{K}{2}\sum_{a=1,2}(\partial_x \partial_y \theta_a+A_{xy})^2
\end{align}

The action is invariant under the following transformation,
\begin{align}
\theta_a\rightarrow\theta_a+f_a(x)+g_a(y),
\end{align}
which are the remnant of the subsystem symmetry of the VPS phase. We can decomposed the two branches as $\theta_{\pm}=\theta_1 \pm \theta_2$.
\begin{align} 
\label{eq:spin}
\mathcal{L}=\frac{K}{2}\sum_{a=+,-}(\partial_t \theta_a)^2-\frac{K}{2}(\partial_x \partial_y \theta_+ +A_{xy})^2-\frac{K}{2}(\partial_x \partial_y \theta_-)^2
\end{align}
We here rescale space and time to set the two coefficients equal.  There is then just one remaining dimensionless paraxmeter as $K=\sqrt{t/u}$. Such a quadratic Lagrangian, in which all terms involve derivatives of the fields, describes a scale invariant phase at long length scales. In a sense it can be viewed as a ``fixed point" Lagrangian. The legitimacy of this approximation and the relevance of compactness will be discussed in detail. The $\theta_+$ mode and the gauge field $A_{xy}$ gap out each other through an analog of the Higgs mechanism. Subsequently, only the $\theta_-$ branch is physical and that is the degree of freedom we will consider here and after.

We first scrutinize the confinement energy between spinons and dipoles in the VPS phase. Due to the subsystem symmetry, the spinons must be excited in quadrupole pairs with two living on the same row/column of the lattice. Following the duality and bosonization argument introduced in Ref.~\cite{xu2007bond,paramekanti2002ring}, we can map the theory in Eq.~\ref{eq:spin} to its dual representation.  As the $\theta_-$ fields are compact with the identification $\theta_-=\theta_-+2\pi Z $, particular types of topological defects will be allowed, which can be most conveniently addressed by passing to a dual representation $N,\phi$ defined on the plaquette centers\footnote{$\hat{n}_-=s_z+1/2$ is the conjugate variable of $\theta_-$.}.
\begin{align} 
\hat{n}_--1/2=\partial_i\partial_j \phi_-,~\hat{N}_-=\partial_i\partial_j \theta_-
\label{dual}
\end{align}
$\hat{N}$ and $\phi$ are a pair of conjugate variables with $\phi$ being discrete-valued and $\hat{N}$ being compact with $N \in [0,2\pi]$. 

The Gaussian part of the dual action is,
\begin{align} 
\mathcal{L}=\frac{1}{2K}(\partial_t \phi_-)^2-\frac{1}{2K}(\partial_x \partial_y\phi_-)^2
\label{dspindual}
\end{align}
In this dual picture, due to the discreteness of $\phi_-$, one can also add vertex operators such as $\cos(4 \pi \partial_i \phi_-)$ in the effective theory. In the VPS phase, these vertex terms are relevant so the spinon and dipole are confined, on which we elaborate further below. In the presence of these vertex operators, the separation of four spinons costs a linear confinement energy proportional to the domain wall length.

We now demonstrate the stability of the algebraic bond liquid phase obtained via melting of the valence plaquette order. In the original representation in Eq.~\ref{eq:spin}, the correlation between two charges $\langle \cos(\theta_-(r)) \cos(\theta_-(0))\rangle$ vanishes at long-wave length due to the subsystem $U(1)$ symmetry. The long-range correlation between Schwinger bosons implies the spinon cannot be excited in pair but instead are created in quartet forms.
The leading order non-vanishing correlation functions are between two dipole operators living on bonds,
\begin{align} 
&\langle \cos(\partial_x\theta_-)(0,0,0) \cos(\partial_x\theta_-)(0,y,\tau) \rangle=\frac{1}{(\tau^2+y^2)^{1/(K \pi^2)}}\nonumber\\
&\langle \cos(\partial_y\theta_-)(0,0,0) \cos(\partial_y\theta_-)(x,0,\tau) \rangle=\frac{1}{(\tau^2+x^2)^{1/(K \pi^2)}}
\end{align}
Notice that the dipole-$i$ correlation function is only nonzero when they are at the same row transverse to the dipole orientation. Thus, the dipoles effectively behave as a $1d$ Luttinger liquid with restricted motion and algebraic correlation on each stripe. This quasi-one-dimensional behavior is crucial for the stability of the bond liquid phase. As the dipole displays $1d$ motion within the same stripe, the quantum fluctuation forbids any dipole condensation with off-diagonal long-range order as a consequence of the Mermin-Wagner theorem. Instead, there appears a quasi-long range order between dipoles akin to the Kosterlitz-Thouless transition.  The quasi-long range order between dipoles corresponds to a spinon-quadrupole excitation with four spins living on the corner of a thin stripe.

To explicate the stability of the bond liquid phase, we need to consider the compactness of the $\theta_-$ term. In the dual representation in Eq.~\ref{dspindual}, the $\phi_-$ is chosen to be half-integers so we consider the vertex operators $V=\cos( 4\pi \partial_i\phi_-),V'=\cos(2\pi\partial^2_i \phi_-)$. When $K$ is small, these vertex operators turn out to be relevant, so the system is driven into a Mott phase which essentially breaks the crystalline symmetry.  In Ref.~\cite{xu2008resonating,paramekanti2002ring}, it was shown that there is a finite region for $K>K_c$ where all vertex operators are irrelevant, so the algebraic bond liquid is stable. Hence, there could potentially appear an algebra bond liquid phase near the VPS phase.

When $K<K_c$, the vertex operator proliferates and thus creates a spin gap. According to the Lieb-Schultz-Mattis theorem, such a gapped spin-1/2 model does not support a featureless Mott phase, so the ground state must break lattice symmetry, which corresponds to the VBS order or VPS order depending on the microscopic symmetry of the vertex operator. For $V_i=\cos( 4\pi \partial_i\phi_-)$, the vertex operator creates a kink for the dipole-i along the transverse stripe. If both $V_x,V_y$ proliferates, the system breaks translation on both directions and thus falls into the stripe order or a plaquette order, depending on the sign of $V_i$ \cite{xu2007bond}. When $\partial_i\phi_-=0$, the system becomes prone to plaquette order as
\begin{align} 
&\cos( 2\pi \partial_i\phi_-)\sim\cos( 2\pi \partial_y\partial_x \theta_-)e^{i\pi r_i},
\end{align}
which exactly agrees with our VPS picture.  
Based on this observation, we conclude that the proliferation/suppression of $V_i$ drives the transition between a VPS phase and the algebraic bond liquid.

When $\partial_i\phi_-=\pi/2$, the system favors a stripe order as
\begin{align} 
&\sin( 2\pi \partial_i\phi_-)\sim(\hat{n}_--1/2)e^{i\pi r_i}
\end{align}
Finally, we mention that, in Ref.~\cite{tay2011possible,tay2010possible}, the authors present a microscopic boson model with various ring-exchange terms which supports a phase transition between plaquette order and the bond liquid phase, with a possible competing order toward a charge density wave. This opens a search for the unconventional plaquette melting quantum phase transition in concrete spin models.

\subsection{Signatures of the algebraic bond liquid phase}

The algebraic bond liquid phase is engendered by the quasi-one-dimensional dipole fluctuations. Due to the restricted motion of spinons, a single spinon excitation is gapped and the spinon correlation $\langle e^{i \theta_a(r)} e^{i \theta_a(r')} \rangle$ vanishes at long wavelength due to the subsystem $U(1)$ symmetry. The bond correlator, denoting the dipole correlation of spinon pairs, has quasi-long-range order as,
\begin{align} 
&\langle e^{i \theta_a(r)} e^{-i \theta_a(r+e_x)}e^{i \theta_a(r+y)} e^{-i \theta_a(r+e_x+y)} \rangle \nonumber\\
&\underrightarrow{y\rightarrow \infty}  =\langle e^{i \partial_x \theta_-(r)} e^{i \partial_x \theta_-(r+y)}  \rangle 
=\frac{1}{(y)^{1/(K \pi^2)}}\nonumber\\
&\langle e^{i \theta_a(r)} e^{-i \theta_a(r+y)}e^{i \theta_a(r+e_x)} e^{-i \theta_a(r+e_y+x)}\nonumber\\
&\underrightarrow{x\rightarrow \infty} =\langle e^{i \partial_y \theta_-(r)} e^{i \partial_y \theta_-(r+x)}  \rangle =\frac{1}{(x)^{1/(K \pi^2)}}
\end{align}
It is worth mentioning that the dipole correlation is anisotropic and only displays algebraic order along the transverse direction, as a consequence of the subdimensional behavior of the dipoles. In particular, the quasi-one-dimensional motion of the dipole, originating from the subsystem charge conservation law, is crucial for the stability of the bond liquid phase.  While spontaneous $U(1)$ symmetry breaking is generally expected in $2d$ quantum systems, a subsystem $U(1)$ symmetry consists of independent symmetry operations acting on an extensively large set of 1-dimensional lines.  The quantum fluctuations thereby suppress dipole long range order, so the Mermin-Wagner theorem still applies.  As a result, the bond correlations decay as a power law due to the absence of spontaneously broken subsystem $U(1)$ symmetry in $2d$.

However, as the spinon is the emergent fractionalized degree of freedom, its four point correlation cannot be measured directly. Instead, we can measure the magnon pair correlator,
\begin{align}
&S^{+}=e^{i (\theta_1-\theta_2)}, S^{-}=e^{-i (\theta_1-\theta_2)}\nonumber\\
&\langle S^{+}(r) S^{-}(r+e_x) S^{+}(r+y) S^{-}(r+e_x+y) \rangle \nonumber\\
&=\langle e^{i \theta_1(r)} e^{-i \theta_1(r+e_x)}e^{i \theta_1(r+y)} e^{-i \theta_1(r+e_x+y)} \rangle \times \nonumber\\
&
\langle e^{i \theta_2(r)} e^{-i \theta_2(r+e_x)}e^{i \theta_2(r+y)} e^{-i \theta_2(r+e_x+y)} \rangle \nonumber\\
&=\frac{1}{(y)^{1/(K \pi^2)}}\nonumber\\
&\langle S^{+}(r) S^{-}(r+e_y) S^{+}(r+x) S^{-}(r+e_y+x) \rangle \nonumber\\
&=\frac{1}{(x)^{1/(K \pi^2)}}
\end{align}
Which also renders an algebraic correlation. The above result is based on the one-loop correction where the magnon correlation is simply the product of two Schwinger boson correlators. Including higher loop corrections can potentially change the exponent of the algebraic correlation.

If we go back to the square lattice structure with unit length $a$, the spinon pair between links can only hop along the transverse direction with even lattice units $2a$ (we will take $a=1$ henceforward). Thus the periodicity of the unit cell is doubled and the Brillouin zone sits in the region $k_i \in [-\pi/2,\pi/2]$ with dispersion as,
\begin{align} 
&\omega(k)=|\sin(k_x)\sin(k_y)| 
\label{ex}
\end{align}
The low-energy dispersion displays a zero-energy nodal line which qualitatively changes the IR behavior, including transport features and entanglement.  Such an excitation spectrum with nodal lines can be measured in terms of the static structure factor for the $S_{z}$ correlator.  In our slave boson representation, the $S_z$ number corresponds to the imbalanced charge density between the two Schwinger bosons as $S_{z}\sim n_1-n_2$, so the static structure factor for the $S_{z}$ correlator can be written in terms of the slave boson correlator as,
\begin{align} 
&S^{zz}(k)\sim \langle (n_1(k)-n_2(k))(n_1(-k)-n_2(-k) )\rangle\nonumber\\
& =\langle (n_1(k)n_1(-k)+n_2(k)n_2(-k) )\rangle\nonumber\\
& \sim |\sin(k_x)\sin(k_y)|
\label{str}
\end{align}
This structure factor can be measured using conventional experimental techniques, such as inelastic neutron scattering and electron spin resonance.

As opposed to the usual bosonic superfluid theory, whose low energy mode condenses at zero momentum, the bond liquid phase contains two nodal lines along the two axes. At each fixed momentum $k_x$, the dispersion resembles a relativistic $1d$ theory, $E=v_f k_y$. Such a bond liquid phase, with low-energy modes containing nodal lines, is termed as a ``Bose surface", in analogy with the $2d$ Fermi liquid with a Fermi surface. For both Fermi surfaces and Bose surfaces, each patch with fixed transverse momentum carries a linearly dispersing $1d$ mode.  Due to the existence of the Bose surface, the low energy transport behavior of the bond liquid phase is qualitatively different from the usual photon gas or weakly interacting bosons.  For each fixed nonzero momentum $k_x$, the low energy modes display linear dispersion with respect to $k_y$, akin to $1d$ relativistic bosons. In particular, due to the nodal lines at $k_x,k_y=0$ with a subextensive number of quasi-1d modes, the specific heat at low temperature scales as\cite{paramekanti2002ring},
\begin{align} 
&C_v \sim T \ln (1/T)
\end{align}
which is similar to the marginal Fermi liquid theory in 2d.

The excitation in Eq.~\ref{ex} corresponds to the dipole excitation containing two spinons between bonds. However, as the spinon is conserved on each line, the dipole excitations must be created in pairs. To be concise, dipoles are created in pairs on the same stripe as a consequence of subsystem symmetry so the spinon appears in quadrupole form. This phenomenon is in close analog to the spinon excitations in 1d spin chains where the magnon excitations can fractionalized into two spinon excitations. Consequently, the spin spectrum function covers a broad continuum whose upper and lower limit corresponds to the parallel and anti-parallel motion of the two spinons. Such continuum in the spectral function distinguishes the spinon excitations with the regular magnons with sharp dispersion.

To seek the collective mode of the dipoles, we calculate the spectral function for $\langle B^{\dagger}_{x} B_{x}\rangle$. The $B_{x}=S^{-}(r)S^{-}(r+e_x)$ operator creates a pair of magnons between an $x$-link. For a non-fractionalized bond liquid\cite{paramekanti2002ring,xu2007bond}, whose dipoles are composed of magnon pairs with integer $S_z$ charge on each row, the spectral function has contributions from the magnon pair excitations with a sharp dispersion relation. In our algebraic bond liquid state, the dipole excitations, created in pairs, carry two spinons between the link with half $S_z$ charge on each row/column.  Such a collective excitation can be interpreted as two dipoles on the same stripe moving along the transverse direction with independent dynamics.

In our slave boson theory, the bond operator $B_{x}=S^{-}(r)S^{-}(r+e_x)=e^{-i \partial_x(\theta_1-\theta_2)}$ creates the $x$-dipole excitation for both $z_1,z_2$ slave particles, each carrying half $S_z$ charge. As these slave dipole pairs are deconfined excitations in the bond liquid phase, each propagates along the stripe with independent motion and the collective excitation corresponds to the combination of the two.  To reach such a collective excitation among spinon-pairs, we calculate the static structure factor for the $\langle B^{\dagger}_{x} B_{x}\rangle=\langle e^{i \partial_x(\theta_1-\theta_2)}e^{-i \partial_x(\theta_1-\theta_2)} \rangle$ correlator,
\begin{align} 
&\langle B^{\dagger}_{x}(Q,\Omega) B_{x}(-Q,-\Omega)\rangle \nonumber\\
&=\int dk d \omega  ~e^{i k^2 G_1(k,\omega)} e^{i ( k+Q)^2 G_2(k+Q,\omega+\Omega)}  \nonumber\\
&G_i(k,\omega)=\langle \theta_i(k,\omega) \theta_i(-k,-\omega) \rangle=\frac{1}{\omega^2-E^2(k)}\nonumber\\
&E(k)=|\sin(k_x)\sin(k_y)|
\label{str}
\end{align}
To extract the collective excitation spectrum, we expand the correlator as, 
\begin{align}
&\langle B^{\dagger}_{x}(Q,\Omega) B_{x}(-Q,-\Omega)\rangle =\sum_{m,n} \frac{1}{m! n!}\Pi_{m,n} \nonumber\\
&\Pi_{m,n}(Q,\Omega)\nonumber\\
&=\int dk d \omega  (k^2 G_1(k,\omega))^n ((k+Q)^2 G_2 (k+Q,\omega+\Omega))^m
\end{align}
The poles in each $\Pi_{m,n}(Q,\Omega)$ correspond to a collective mode.  We start with the leading order expansion $\Pi_{1,1}(Q,\Omega)$,
\begin{align}
&\Pi_{1,1}(Q,\Omega)=\int dk \frac{(k^2)( k+Q)^2}{E(k)}\nonumber\\
&[\frac{1}{(E(k)+\Omega)^2-E^2(k+Q)}
+\frac{1}{(E(k)-\Omega)^2-E^2(k-Q)}]
\end{align}

This propagator renders a series of poles as $\Omega(Q)= E(Q-k)+ E(k)$. The energy spectrum at fixed momentum $Q$ covers a broader range depending on the choices of $k$. This continuum spectrum can be understood as the two-dipole excitation with dispersion $E(Q-k)$ and $E(k)$. The total momentum is fixed as $Q$ with $Q-2k$ being relative momentum due to the independent motion of the dipole pair.

\begin{figure}[h]
  \centering
      \includegraphics[width=0.35\textwidth]{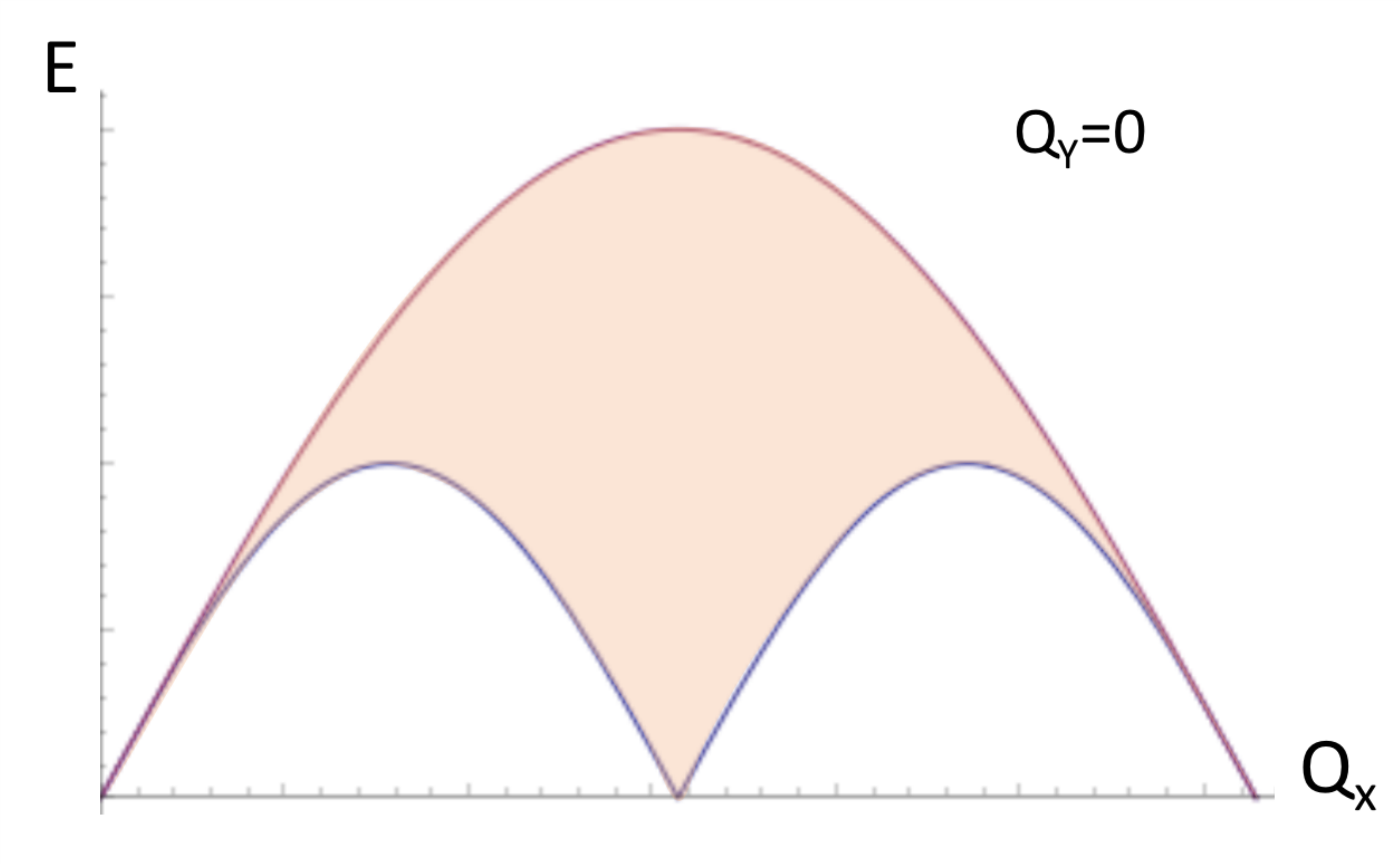}
  \caption{Continuous energy spectrum with respect to $Q_x$ with fixed $Q_y=0$. 
} 
\label{spec}
\end{figure}

In particular, if we fix a slice of the momentum space by taking $Q_y=0,\pi$, the energy spectrum has an upper and lower limit as,
\begin{align}
&\Omega(Q_x,Q_y=0)\sim |\sin(k_x)|+|\sin(Q_x-k_x)|\nonumber\\
&\Omega^{upper}(Q_x)\sim \sin(Q_x/2), ~\Omega^{lower}(Q_x)\sim \sin(Q_x)
\end{align}
The upper or lower limit of the spectrum denotes the parallel or opposite motion for the two dipoles. This collective excitation fills the continuum region between these two limits, as depicted in Fig.~\ref{spec}, which resembles the spinon spectrum in $1d$ anti-ferromagnetic spin chains. Actually, when fixing $Q_y$, we focus on the collective motions of $y$-dipoles along the $x$-direction. As the dipole's motion is restricted along the $x$-stripe, the collective mode is attributed by the dispersion of two dipoles along the stripe\cite{kohno2007spinons}, which can be regarded as the descendent of the spinon in $1d$ spin chains.

Another signature for the algebraic bond liquid is the violation of the area law for entanglement entropy, arising as a consequence of the bose surface. In Ref.~\cite{boseentanglement}, it was demonstrated that if we equally bipartition the bond liquid state on a lattice along the $x$-axis, the entanglement entropy scales as $L_x \ln (L_y)$, which resembles the entanglement entropy for the $2d$ Fermi surface \cite{wolf,widom,swingle}.  Such a violation of the area law can be roughly understood by dividing the Bose surface into small patches over which the surface looks approximately flat, such that each patch can be regarded as a one-dimensional relativistic boson whose entanglement entropy scales as $\ln (L)$.  Summing over the contributions from all patches, the total entanglement entropy should behave as $L\ln (L)$. Such long-range entanglement is smoking-gun evidence for the Bose surface which can be detected in numerical simulations.


\subsection{$SU(3)$ Plaquette defect on triangle lattice }

Our previous discussion based on plaquette order on square lattices can be straightforwardly applied to other bipartite lattices, such as the honeycomb lattice.  In particular, the spinon living in the $C_3$ plaquette defect obeys a conservation law on each $\theta=2\pi/3$ line, so the spinon is also a fracton. A spinon pair, regarded as a dipole, can hop on transverse zigzag stripes as a $1d$ subdimensional particle.

\begin{figure}[h]
  \centering
      \includegraphics[width=0.4\textwidth]{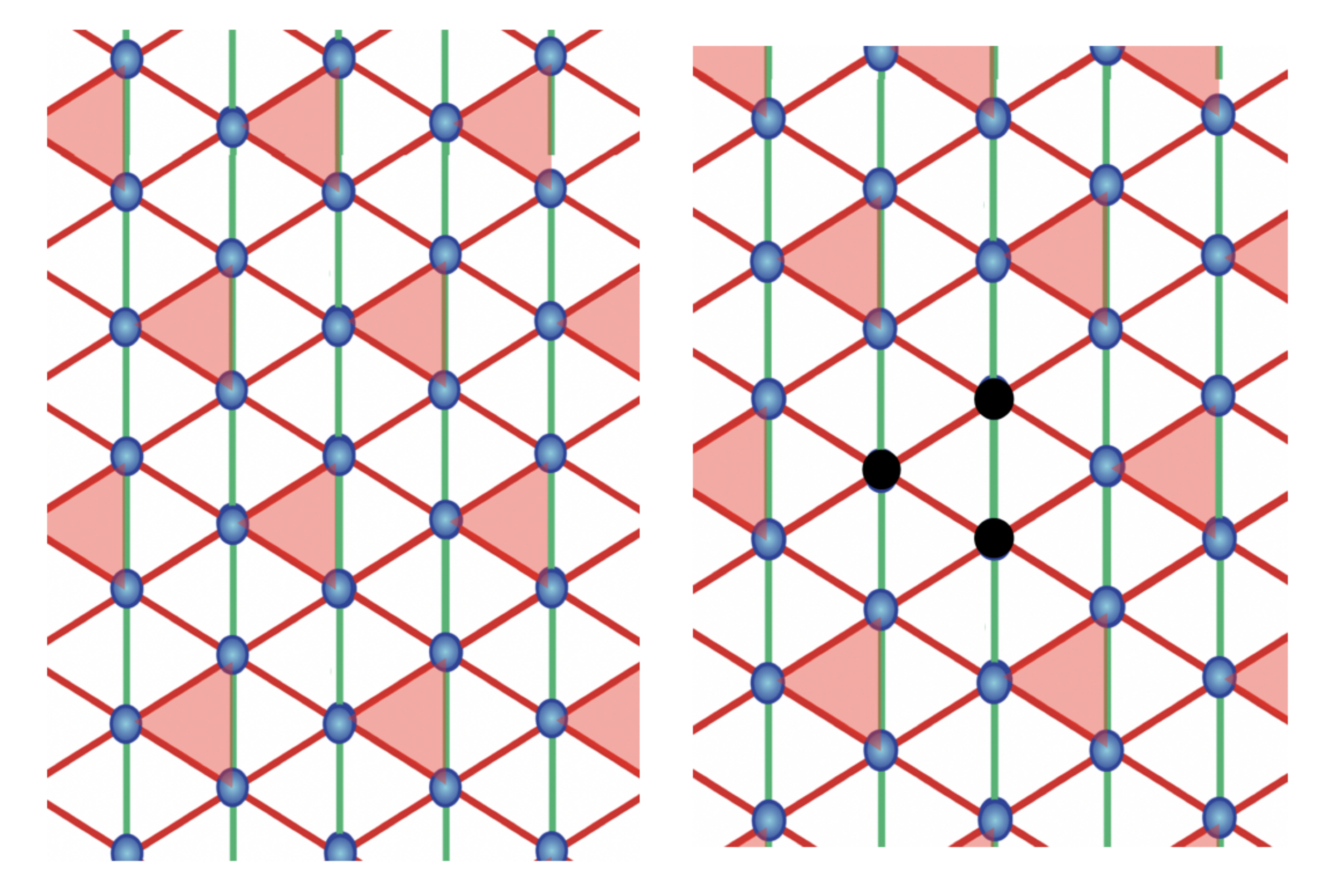}
  \caption{L: $SU(3)$ Plaquette order on the triangular lattice. R: Breaking a plaquette creates three $SU(3)$ spins.
} 
\label{frac}
\end{figure}

Finally, we would like to bring attention to the $SU(3)$ plaquette order on a triangular lattice with possible fractal dynamics. A typical plaquette order contains $SU(3)$ singlets between three $SU(3)$ spins living on the three sites of left-oriented triangles, as illustrated in Fig.~\ref{frac}.  As the $SU(3)$ valence plaquette only lives on the left-oriented triangles, the system breaks $C_3$ rotation symmetry and the ground state contains three VPS configurations related by $C_3$ rotation.

A typical topological defect renders the breaking of $SU(3)$ singlet on one triangle which creates three dangling $SU(3)$ spins. It is obvious that one can not move or separate a single $SU(3)$ spin out of the defect due to fractal conservation laws.  In particular, if we define $P=\pm$ as the plaquette occupancy/vacancy for each left-oriented triangle, the charge conservation law for $SU(3)$ spin is
\begin{align} 
P_{i+e_1} P_{i+e_2} P_{i+e_3}= (-1)^{q^s}
\end{align}
$e_i$ are the three vectors from a site toward the left-oriented triangles, $q_s$ is the $SU(3)$ charge.
Apparently, the $SU(3)$ charge is conserved in each fractal manifold with the shape of a sierpinski triangle. Thus, it is impossible to move a single $SU(3)$ toward any direction due to the special fractal conservation law. Due to the fractal dynamics and restricted mobility of $SU(3)$ spins, the melting of the plaquette order does not engender any deconfined quantum crticial point toward the spin $SU(3)$ breaking state. At this stage, we are still agnostic about the possible phase diagrams near this plaquette order phase, so we leave this as a topic for future study.

\section{Fractons in 3D cube ordered and valence plaquette phases}
\label{sec:3d}

In this section, we extend the scope of our analysis to a $3d$ cubic lattice. We consider properties of the topological defects in $3d$ valence cube solid (VCS) and valence plaquette solid (VPS) phases. We show that the topological defects of these orders are generically fractons. Then we consider possible outcome of the melting transitions of these orders indicated by the fracton dynamics.

\subsection{3D valence cube solid (VCS) order}

A natural generalization of the 2d plaquette order in 3d is the cube order. The cube order for an $SU(2)$ spin-$1/2$ model on a cubic lattice is a state which has resonating clusters of 8 spins on every other cube in the lattice. One can also imagine similar state for models with larger spin symmetry such as $SU(4)$ and $SU(8)$. We argue that topological defects of these cube ordered phases are emergent fractons just as in the 2d plaquette order. We also propose possible neighboring phases assuming the fractonic dynamics persists across the phase transition. In the following, we restrict ourselves to the case of spin-$1/2$ systems on cubic lattices. 

The fractonic nature of the topological defects becomes clear after mapping the cube order to a rank-3 tensor gauge theory. Since cubic lattice is a bi-partite lattice, on each cube, one can define an ``electric" field $E_{xyz}(\bm{r})=(-1)^{i_{\bm{r}}}C(\bm{r})$, where $i_{\bm{r}}=$odd/even if $\bm{r}\in A/B$ sublattice, and $C(\bm{r})=1$ (or $0$) denotes that there is (or not) a resonating cluster on cube $\bm{r}$\footnote{The cube $\bm{r}$ is the adjacent cube of site $\bm{r}$ along the $(111)$ direction.}. $E_{xyz}$ furnishes a rank-3 hollow gauge theory in $3d$. In the low energy Hilbert space, the electric field satisfies the following Gauss's law around a site on the lattice,
\begin{equation}
    \partial_x\partial_y\partial_z E_{xyz}(\bm{r})=(-1)^{i_{\bm{r}}}(1-q(\bm{r})),
    \label{eq:3dGL}
\end{equation}
where $q=0/1$ denotes the number of free spinon at site $\bm{r}$, and all the derivatives should be treated as lattice derivatives. The $q=1$ state corresponds to a topological defect of the cube order, which maps to a point charge of the rank-3 gauge field. 

One can introduce the conjugate field $A_{xyz}(\bm{r})$ for $E_{xyz}(\bm{r})$, namely $[A_{xyz}(\bm{r}),E_{xyz}(\bm{r'})]=\frac{i}{2\pi}\delta_{\bm{r},\bm{r'}}$. The $e^{\pm i A_{xyz}(\bm{r})}$ is the creation/annihilation operator for $E_{xyz}(\bm{r})$ on cube $\bm{r}$. The Gauss law in the low energy subspace in Eq. \ref{eq:3dGL} implies the following gauge transformation for $A_{xyz}$,
\begin{equation}
    A_{xyz}\rightarrow A_{xyz}+\partial_x\partial_y\partial_z\lambda.
\end{equation}
With this gauge transformation, we can locally remove the gauge field. Therefore, there is no local ``magnetic" flux in this rank-3 gauge theory. Physically, this means there is no local resonating process for the cube order. Namely any local adjustments of the cube order parameter inevitably break the Gauss's law constraint. Of course, there are still global flux operators, which can adjust the cube order either on a whole plane or along a straight line, analogous to the global flux operators appearing in the 2d case in Eq. \ref{eq:Gflux}.

 Since the topological defect, which traps a single spinon, appears as the matter field that couples to the rank-3 gauge field, it is a fracton that cannot move along any direction.  Specifically, the Gauss's law of this theory implies conservation of all components of dipole and quadrupole moments, along with certain components of the octupole moment.  The mobility of other point defects of the cube order are also easy to determine. From the physical picture of cube order, one can see that the spinon dipoles are also immobile. While the spinon quadrupole on a plane is movable along the normal direction, hence it is a linenon.
 
 Now we consider possible melting transitions if this fractonic constraint is kept all the way through. Since the spinon monopoles and dipoles have no mobility at all, it is hard to consider their condensation. The most probable way to drive the system out of the cube order is to proliferate the planar spinon quadrupole, which is movable along different lines. However, $1$-dimension cannot host true long range order of continuous symmetry. Therefore, the resultant phase may be an algebraic spin liquid similar as the $2d$ case. 

 Let us write down the low energy field theory which encodes the coupling between the rank-3 gauge field and spinon matter field,
\begin{align}
\nonumber
     \mathcal{H}=&U\sum_{\bm{r}}E_{xyz}(E_{xyz}-(-1)^{i_{\bm{r}}}) \\ \nonumber
   &+u(\sum_{a=1,2}n_a-1)^2 +t\sum_{a=1,2}\cos(\partial_x\partial_y\partial_z\theta_a+A_{xyz})
\end{align}
where we have again used the $CP^1$ map as in Eq. \ref{eq:cp1} to fractionalize the spin at the topological defects. As before, the $n_1$ and $n_2$ are the number operators of the two bosonic spinons. We adopt a roton approximation for the spinons. Correspondingly, the $\theta_1$ and $\theta_2$ are the phases of the two bosons. This theory is invariant under the following symmetry transformation,
\begin{equation}
    \theta_a\rightarrow \theta_a+g_{1a}(x,y)+g_{2a}(y,z)+g_{3a}(z,x).
    \label{eq:sym}
\end{equation}

Consider an easy plane limit, namely adding a term $u'\sum_{a=1,2}(n_a-\frac{1}{2})^2$ to favor $\bar{n}_1=\bar{n}_2=\frac{1}{2}$. With a large $t$, the system tends to fall into an ordered state of $\theta_1$ and $\theta_2$. In a spin-wave approximation, we can expand the cosine term and take the gaussian theory. In the resultant theory, the gauge fields and the $\partial_x\partial_y\partial_z(\theta_1+\theta_2)$ mode gap out each other through an analog of the Higgs mechanism. The only physical mode left is $\partial_x\partial_y\partial_z(\theta_1-\theta_2)$. Let us label $\theta=\theta_1-\theta_2$, and $n=n_1-n_2$. The continuum theory now reads
\begin{align}
     \mathcal{H}_G=\frac{1}{8K}n^2+\frac{K}{2}(\partial_x\partial_y\partial_z\theta)^2,
\end{align}
where we have rescaled spatial coordinates to simplify the Hamiltonian to this form with $K\sim \sqrt{t/u'}$. Because of the symmetry in Eq. \ref{eq:sym}, there is no cosine terms for $\theta$. The gaussian theory describes a $3d$ generalization of the $2d$ algebraic liquid phase, in which the planar quadrupole operators acquire algebraic correlations. For example, the quadrupole operators $Q_z=\cos(\partial_x\partial_y\theta)$ have zero equal time correlation along $x$ and $y$ direction due to the symmetry in Eq. \ref{eq:sym}, however, power-law correlation along $z$-direction,
\begin{equation}
\langle Q_z(0,0,z)Q_z(0,0,0)\rangle\sim \frac{1}{|z|^{1/(\pi^2K)}}.
\end{equation}
Planar quadrupole operators along different planes have similar power law correlations. The directional power-law correlations of the planar quadrupoles are the remnant of their fracton dynamics in VCS phase. 

In the gaussian theory, we have ignored the compactness of the $\theta$ variable. To justify the stability of the gaussian theory, one has to consider the relevancy of the vertex operators. To do this, it is most convenient to go to the dual description. We can have a duality map 
\begin{equation}
      \frac{1}{2}n=\partial_x\partial_y\partial_z\phi,\ \ \ \partial_x\partial_y\partial_z\theta=N  
\end{equation}
where $\phi$ and $N$ are conjugate variables defined on the dual lattice sites. $\phi$ should take values in $\mathbb{Z}/2$, while $N\in [0,2\pi)$. The dual hamiltonian reads
\begin{align}
     \mathcal{H}_D=\frac{K}{2}N^2+\frac{1}{2K}(\partial_x\partial_y\partial_z\phi)^2+....
\end{align}
In this continuous field theory, we have regarded $\phi$ fields as real number. However, since the $\phi$ variable actually takes values in $\mathbb{Z}/2$, we have to include cosine terms, namely the vertex operators, to reinstall the integral constraint of the $\phi$ fields. There are various vertex operators that can appear in the theory, for example, $V=\cos(4\pi\phi)$, $V_x=\cos(4\pi\partial_x\phi)$, $V_{xy}=\cos(4\pi\partial_x\partial_y\phi)$ and so on. The task is to determine the scaling dimensions of these vertex operators and to inspect if they are relevant at the gaussian fixed point. The gaussian theory has an emergent symmetry,
\begin{equation}
    \phi(x,y,z)\rightarrow \phi(x,y,z)+f_1(x,y)+f_2(y,z)+f_3(z,x)
\end{equation}
which restricts the correlations of the vertex operators. The correlations between $V$ at different points are zero, similarly for $V_x$, due to the emergent symmetry. The $V_{xy}$ operator can have non-zero correlation functions only along $z$ direction. The most relevant vertex operator has correlation function as the following,
\begin{equation}
    \langle V_{xy}(0,0,z)V_{xy}(0,0,0)\rangle \sim\frac{1}{|z|^{\eta'K}},
\end{equation}
with $\eta'>0$ depending on the UV definition of the vertex operator. We can see that for large enough coupling $K$, the vertex operators in the theory can be all irrelevant, hence, the algebraic spin liquid is stable. 

\subsection{Tensor gauge theory and fractons in 3D VPS}

Consider an $SU(2)$ spin model on a cubic lattice. In analogy with the previously discussed $2d$ valence plaquette order, the $3d$ plaquette order contains 12 distinct plaquette order patterns, as there are 12 plaquettes adjacent to a single site. Such plaquette order breaks cubic symmetries and translations while maintaining the $SU(2)$ spin rotation symmetry. We now show the valence plaquette order on the cubic lattice, similar to the $2d$ case, can be mapped to a hollow rank-2 gauge theory \cite{xu2008resonating,bulmash2018higgs,ma2018higher}.

\begin{figure}[h]
  \centering
      \includegraphics[width=0.35\textwidth]{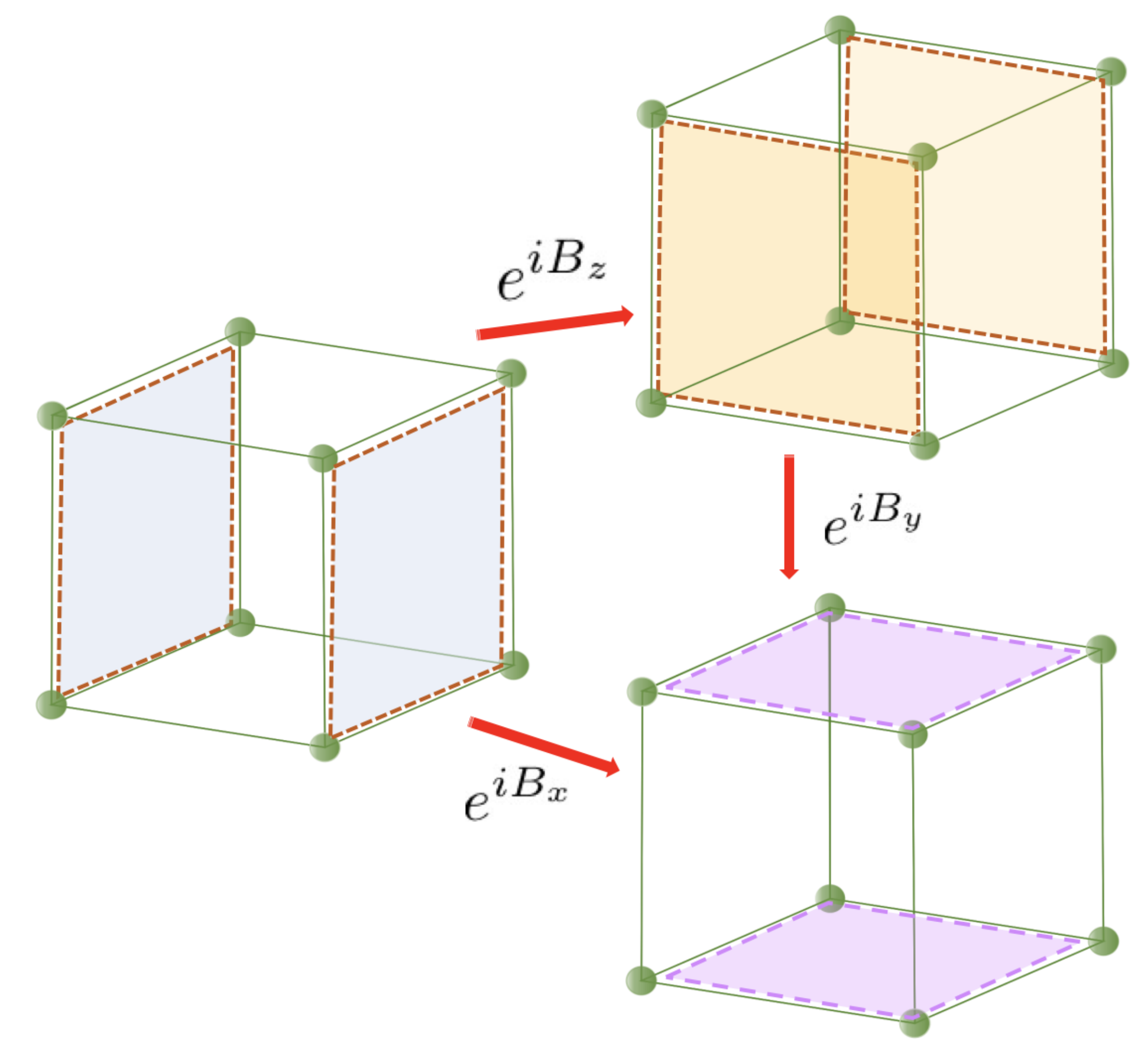}
  \caption{The flux operator create resonance between different plaquette configurations.
} 
\label{3d}
\end{figure}

We denote the plaquette order on each $i$-$j$ square as a tensor electric field,
\begin{align} 
E_{ij}(\bm{r})=(-1)^{i_{\bm{r}}} P_{ij},
\end{align}
where the binary variable $P_{ij}=1\,(0)$ corresponds to the valence plaquette occupancy (vacancy) on each square. $i_{\bm{r}}$ is again the sublattice index for site $\bm{r}$. As the plaquette lives on three $i$-$j$ planes, there are three components of the tensor electric field $E_{xy},E_{zy},E_{xz}$, forming a symmetric rank-2 hollow ($i.e.$ purely off-diagonal) gauge theory \cite{xu2008resonating,bulmash2018higgs,ma2018higher}.  The gauge field $A_{ij}$, as the conjugate variable of $E_{ij}$ satisfying the usual commutation relation $[A_{ij},E_{ij}]=\frac{i}{2\pi}$, is the operator which creates/annihilates a valence plaquette on each square and thus enables plaquette fluctuation. This is different from the $2d$ case as in $2d$ the plaquette order cannot fluctuate locally. Since each spin on a site is only entangled with one of the 12 adjacent plaquette clusters, one can write down the analogy of the Gauss's Law for the rank-2 gauge field as the following,
\begin{align} 
\partial_i \partial_j E_{ij}(\bm{r})=(-1)^{i_{\bm{r}}} (1-q^s(\bm{r})),
\label{gauss3d}
\end{align}
where $\partial_i$ should be regarded as the lattice derivative on the cubic lattice and the Gauss's law respects the cubic symmetry.  This Gauss's law exactly resembles the hollow rank-2 symmetric gauge theory in 3D as the $U(1)$ generalization of the X-cube fracton model \cite{xu2008resonating,bulmash2018higgs,ma2018higher,Vijay2015-jj,Vijay2016-dr}.  Due to the particular double derivative in Eq.~\ref{gauss3d}, the spinon is conserved on each $i$-$j$ plane, so the theory respects a subsystem planar $U(1)$ symmetry. Consequently, a fundamental topological defect, which carries a single spinon, is a fracton that is restricted from moving in any direction.  In addition, the topological defects which host a pair of spinons along a link can hop within the $2d$ plane which is perpendicular to its dipole orientation.  As opposed to the $2d$ VPS order, the plaquette configuration on the cubic lattice can fluctuate and resonate locally on each cube. These local fluctuations defined on each cube can be mapped to three types of magnetic flux operators in the gauge theory language,
\begin{align} 
B^a=\epsilon^{ija}\partial_i A_{ja},
\end{align}
which are invariant under the following gauge transformation,
\begin{align}
A_{ij}  \rightarrow A_{ij}+\partial_i \partial_j. \alpha
\end{align}
Different from the vector $U(1)$ gauge theory in $3d$, the flux operators here are point-like excitations obeying the identity $\sum_{a}B^a=0$. In particular, the flux excitations are also fractons, known as lineons\cite{Vijay2016-dr,xu2008resonating}, which only move along straight lines. Such flux operators flip the plaquette configuration on one cube, as in Fig.~\ref{3d}, which generates resonant states between different flippable plaquette configurations.
 
A typical hamiltonian for the pure compact rank-2 gauge theory can be written as the following,
\begin{align} 
\nonumber
\mathcal{H}=&  U\sum_{\bm{r}}\sum_{i,j} E_{ij}(E_{ij}-(-1)^{i_{\bm{r}}})+V\sum_{\bm{r}}(\epsilon_{kij}\partial_kE_{ij})^2\\
&-T
\sum_{\bm{r}}\sum_{a=1,2,3}\cos(B^a),
\label{em}
\end{align}
the electric fields are subject to the Gauss's law on every site. Due to the proliferation of topological defects, namely the $2\pi$ instanton tunneling which turns out to be relevant, the pure rank-2 gauge theory is generically in a confined phase with crystalline orders\cite{xu2008resonating}.

Next we consider matter fields and possible Higgs-like transitions in this rank-2 gauge theory. The matter field that couples to the gauge theory are the topological defects of the VPS order. 
\begin{align}
    \mathcal{H}_{m}=u\sum_{\bm{r}}(\sum_{a=1,2}n_a-1)^2+t\cos(\partial_i\partial_j\theta_a+A_{ij}),
\end{align}
where again we have use the $CP^1$ formalism to represent the spinon trapped in the topological defects. The single spinon are fractons that cannot move, while a pair of spinon can move in the plane that is perpendicular to its dipole moment. Let us use this hierarchy of matter field mobilities to infer the possible nearby phases. Due to the restricted motion of the spinon, a direct condensation of spinons is inhibited. The leading ordering instability should be dipole condensation, where the spinon pair between links acquires coherence along a 2d plane. Such condensation restores the mobility of the spinon along the dipole direction and thus breaks subsystem symmetry. Depending on the microscopic Hamiltonian, the spinon pair condensate could engender a valence bond solid or liquid phase. In the following, we will consider a case with strong anisotropy where the melting transition leads the system to a valence bond solid state. 

\subsection{An anisotropic VPS}

We now consider a special limit of the valence plaquette solid with strong anisotropy along the $z$-direction, namely the valence plaquettes energetically favor to be on the $xz, yz$ squares. In this case, one can imagine a columnar plaquette ordered ground state along $xz$ or $yz$ direction, which spontaneously breaks the $C_4$ rotation along $z$ direction. In this anisotropic case, the low energy Hilbert space has $E_{xy}=0$. Thus the Gauss's law is reduced to,
\begin{align} 
\partial_x \partial_z E_{xz}(\bm{r})+\partial_y \partial_z E_{yz}(\bm{r})=(-1)^{i_{\bm{r}}} (1-q^s(\bm{r})).
\end{align}
Notice there are only two electric operators, corresponding to the plaquettes on $xz$ and $yz$ planes. A single spinon is still immobile, while a pair of spinons along a $z$-link can hop on the $xy$-plane as a $2d$ subdimensional particle. A spinon pair along an $x$ or $y$ link is conserved in each $z$-stripe, so they are restricted to fluctuate along $z$.

The valence plaquette configuration can fluctuate and resonate between the side faces of the cube. Such a local plaquette configuration resonance corresponds to the flux operator,
\begin{align} 
B^z=\epsilon^{ijz}\partial_i A_{jz},\quad (i,j \in x,y)
\end{align}
which is invariant under the following gauge transformation,
\begin{equation}
A_{iz}  \rightarrow A_{iz}+\partial_i \partial_z \alpha
\end{equation}
The effective field theory description for this anisotropic limit can be written as 
\begin{align} 
\nonumber
\mathcal{H}=&U\sum_{\bm{r}}(E_{xz}(E_{xz}-(-1)^{i_{\bm{r}}})+E_{yz}(E_{yz}-(-1)^{i_{\bm{r}}})) \\
&+V\sum_{\bm{r}}((\partial_yE_{xz})^2+(\partial_xE_{yz})^2)-T\cos(B^z),
\end{align}
 which is derived from Eq.~\ref{em} by enforcing $E_{xy}=0$ and confining the $A_{xy}$ gauge field component.

\begin{figure}[h]
  \centering
      \includegraphics[width=0.45\textwidth]{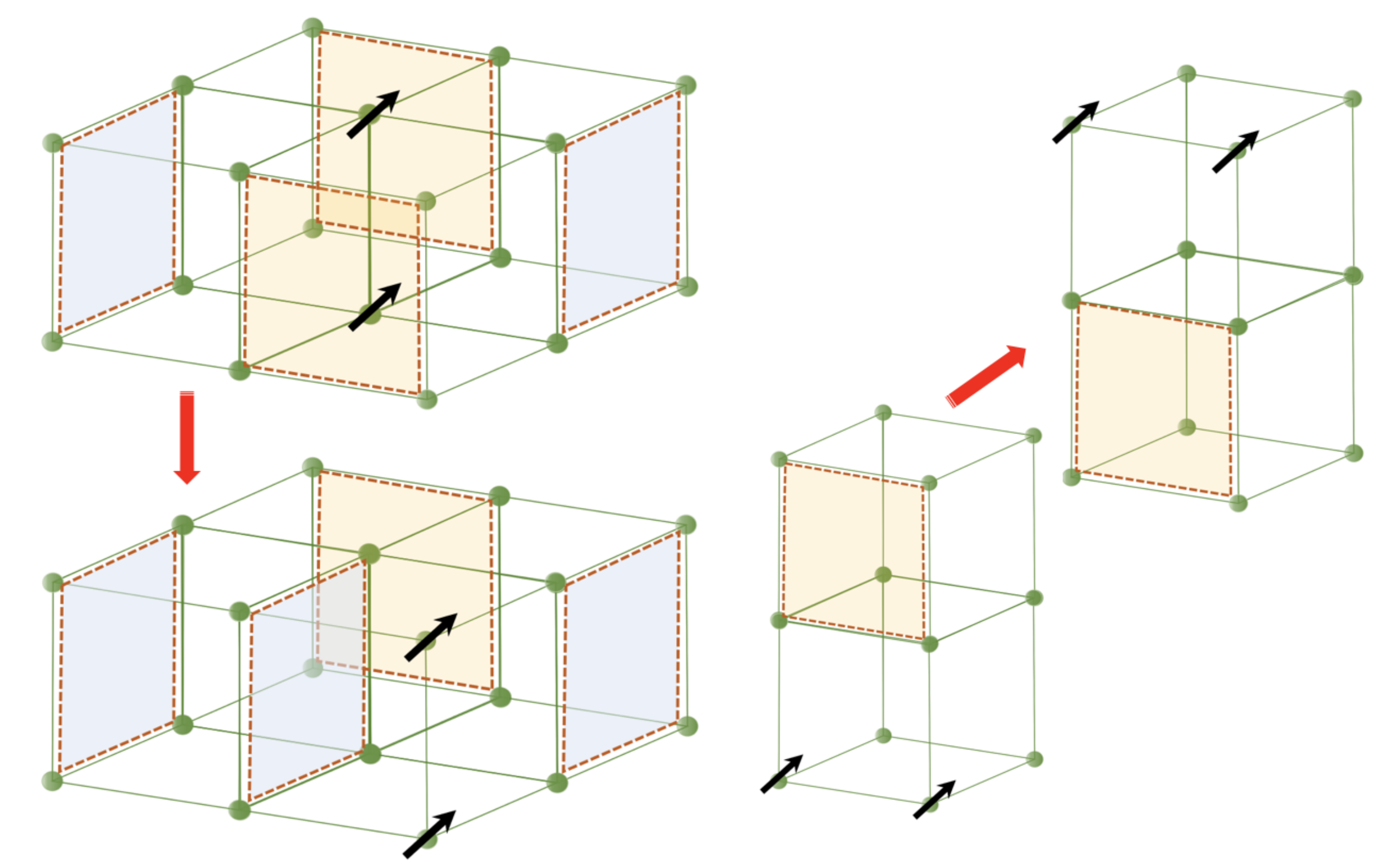}
  \caption{A pair of spinon can hop on the 2D plane perpendicular to its polarization
} 
\label{c4}
\end{figure}

As the plaquette order fluctuates, there can appear vortex configurations where the four plaquette patterns related by $C^z_4$ symmetry meet at a point, forming a $2d$ vortex, as shown in Fig.~\ref{c4}.
This $C^z_4$ vortex defect, carrying a spinon pair along a $z$-link, can fluctuate on the $xy$-plane. Similarly, there are spinon pairs along $x$ and $y$-link. However, they can only fluctuate along $z$-direction due to the fracton constraint and anisotropy. 

Now let us consider a zero temperature quantum melting transition driven by condensation of these spinon pairs. We start from a particular VPS state, say all the plaquette is along $xz$ plane. The melting transition will try to break the plaquettes. We can imagine a situation where the plaquette can only break into two $z$-direction dimers in the hilbert space. In this limit, we can imagine a VPS to VBS phase transition. Due to the fractonic nature of the dimers, two things need to be accomplished during the transition. a) In each double-layer, the $z$-dimer proliferates and destroys the plaquette order within a layer. b) The layers establish coherence. If a) happens before b), then we have a transition that features dimensional decoupling. If b) happens first, we have a transition that has reduced effective dimensions. 

Let us first analyze the phase transition within a double layer. Due to the anisotropy, the resonating plaquettes are formed between the two layers. If we view the system from the top, the VPS order actually is projected to a 2d VBS pattern. In this top view, the interlayer singlet pair is the vortex core of the VBS pattern. Therefore, for the double layer system, the VPS to VBS transition is mapped to a VBS melting transition. The key difference between this transition and the 2d DQCP is that the VBS vortices here do not have spinon degree of freedom. The low energy field theory for such a transition is the following
\begin{equation}
\mathcal{L}_{2d}=|(\partial_\mu-i a_\mu)\phi|^2+r|\phi|^2+g|\phi|^4+\frac{1}{4e^2}f^2+...,   
\end{equation}
which is similar to the theory of DQCP as in Eq. \ref{eq:DQCP} except the matter field $\phi$ here is a single scalar. We still need to include the 4-fold instantons in the theory, which is allowed by translation symmetry. In the particle-vortex dual picture, the transition can also be described as a 3d XY transition with four-fold anisotropy. It is known that the 4-fold anisotropy is irrelevant at the 3d XY transition. Thus, the VPS to VBS transition in the double layer system is in the 3d XY universality class. 

Now we consider the coupling between these 2-dimensional critical theories along the $z$ direction. Notice that for each bilayer there is an independent emergent gauge field, which constrains the possible coupling between layers. The lowest order gauge invariant coupling of the matter field between layers is
\begin{equation}
\mathcal{L}_c=\sum_{ij}\lambda_{ij}|\phi_i|^2|\phi_j|^2,
\end{equation}
where $i,j$ are layer indices. This coupling is irrelevant under RG at the 3d XY fixed point. However, we also need to include the monopole tunneling between layers. 
\begin{equation}
\mathcal{L}_m=\sum_{ij}g_{ij}(\mathcal{M}^\dagger_i\mathcal{M}_j+h.c.)
\end{equation}
This monopole tunneling term is gauge invariant. More importantly, this term is highly relevant in the 2d critical theory. This term will lock the gauge field between different layers. In the end, there is only one gauge field, which has $2d$ characters, across the $3d$ system. This suggest that the layers should first establish coherence and then go through the VPS-VBS transition all together. Physically, this means that the plaquette patterns of different layers along $z$ directions synchronize. The transition are driven by proliferation of straight vortex lines along $z$-directions. This suggests that the VPS-VBS transition looks like a $2d$ transition despite the system is 3-dimensional. We have a transition that features a reduced effective dimension.

\section{Summary}\label{sec: summary}

In this work, we have shown how fracton physics gives important insight into melting transitions of valence plaquette solids.  The topological defects of this type of spatial order are characterized by fractonic mobility constraints, as elucidated by a mapping onto a symmetric tensor gauge theory.  Specifically, the individual vortices are completely immobile, while dipoles of vortices exhibit 1-dimensional behavior.  This restricted mobility can prevent a direct condensation of single vortices, precluding a continuous transition from the VPS phase to a simple N\'eel state.  Rather, a continuous melting transition of a VPS tends to involve condensation of vortex dipoles, giving rise to an intermediate phase between the VPS and N\'eel state.  This intermediate phase can take different forms depending on the microscopic details of the dipoles, such as various types of bond order.  A particularly interesting possibility in two dimensions is an algebraic bond liquid, which serves as a stable intermediate gapless phase, in agreement with numerics on certain $2d$ Heisenberg models \cite{sheng}.  We have discussed several signatures of this algebraic bond liquid, such as its structure factor, specific heat, and entanglement properties.  We have also discussed the extension of these ideas to three-dimensional valence plaquette and valence cube solid phases.  We show that, a particular anisotropic type of plaquette order can undergo a continuous transition to a bond-ordered phase via a quantum critical point that features a reduced effective dimension.

\begin{acknowledgments}
We thank Cenke Xu, Meng Cheng, Chong Wang and T. Senthil for helpful discussion. YY is supported by PCTS fellowship at Princeton University. ZB is supported by the Pappalardo fellowship at MIT. This work was performed in part at Aspen Center for Physics, which is supported by National Science Foundation grant PHY-1607611.
\end{acknowledgments}

\section*{Appendix: Parton view of the spinon bond liquid}

In Ref.~\cite{tay2011possible,tay2010possible}, the authors present a parton construction for the bond liquid phase, which captures several salient features of this phase including subdimensional motion and long-range entanglement. Here we apply their parton perspective to illustrate the emergence of subdimensional particle.

Writing the boson operator as $b_a^{\dagger}=v_a^{\dagger}h_a^{\dagger}$. The Schwinger boson operator is further fractionalized into the vertical and horizontal bosons $v_a^{\dagger},h_a^{\dagger}$. At the mean-field level, $\langle v^{\dagger}_iv_{i+e_x}\rangle=\langle h^{\dagger}_ih_{i+e_y}\rangle \neq 0$, as each vertical or horizontal parton only hops along the $x$ or $y$ direction, as dictated by the following Hamiltonian,
\begin{align} 
&\mathcal{H}_{v,h}=|(\partial_x+a_x)v|^2+|(\partial_y-a_y)h|^2+\cos(\nabla \times a)+\frac{e^2}{g}
\end{align}
Each parton $v^{\dagger},h^{\dagger}$ is analogous to a 1d relativistic boson coupled with an emergent gauge field $a$. The strong fluctuation of the gauge field induces strong interaction between the two partons and projects the parton state to the physical Hilbert space. The vertical parton carries gauge charge $a$, so its algebraic correlation along $x$ should be modified by the gauge fluctuation. Besides, a pair of vertical parton is charge neutral with power-law correlation $\langle v^{\dagger}_iv_{i+e_y}v_{i+x} v^{\dagger}_{i+x+e_y} \rangle=1/|x|^{\eta}$. At the mean field level, this parton pair correlation exactly corresponds to the bond correlation,
\begin{align} 
&\langle b^{\dagger}_i b_{i+e_y}b_{i+x} b^{\dagger}_{i+x+e_y} \rangle \nonumber\\
&=\langle v^{\dagger}_iv_{i+e_y}v_{i+x} v^{\dagger}_{i+x+e_y} h^{\dagger}_i h_{i+e_y}h_{i+x} h^{\dagger}_{i+x+e_y} \rangle \nonumber\\
&=\langle v^{\dagger}_i v_{i+e_y} v_{i+x} v^{\dagger}_{i+x+e_y}\rangle \langle  h^{\dagger}_i h_{i+e_y}\rangle \langle h_{i+x} h^{\dagger}_{i+x+e_y} \rangle \nonumber\\
&=c \langle v^{\dagger}_i v_{i+e_y} v_{i+x} v^{\dagger}_{i+x+e_y}\rangle
\end{align}
This parton construction provides a pictorial understanding of the bond liquid phase, which has low-energy behavior, including entanglement entropy, very similar to that of $1d$ relativistic bosons. In the mean field level, each parton forms a 1d spin chain along horizontal/vertical with elementary spinon excitation. After parton projection, the vertical/horizontal spinon motions are bound together as a spinon pair between a link hopping along the transverse direction.


\begin{thebibliography}{88}%
\makeatletter
\providecommand \@ifxundefined [1]{%
 \@ifx{#1\undefined}
}%
\providecommand \@ifnum [1]{%
 \ifnum #1\expandafter \@firstoftwo
 \else \expandafter \@secondoftwo
 \fi
}%
\providecommand \@ifx [1]{%
 \ifx #1\expandafter \@firstoftwo
 \else \expandafter \@secondoftwo
 \fi
}%
\providecommand \natexlab [1]{#1}%
\providecommand \enquote  [1]{``#1''}%
\providecommand \bibnamefont  [1]{#1}%
\providecommand \bibfnamefont [1]{#1}%
\providecommand \citenamefont [1]{#1}%
\providecommand \href@noop [0]{\@secondoftwo}%
\providecommand \href [0]{\begingroup \@sanitize@url \@href}%
\providecommand \@href[1]{\@@startlink{#1}\@@href}%
\providecommand \@@href[1]{\endgroup#1\@@endlink}%
\providecommand \@sanitize@url [0]{\catcode `\\12\catcode `\$12\catcode
  `\&12\catcode `\#12\catcode `\^12\catcode `\_12\catcode `\%12\relax}%
\providecommand \@@startlink[1]{}%
\providecommand \@@endlink[0]{}%
\providecommand \url  [0]{\begingroup\@sanitize@url \@url }%
\providecommand \@url [1]{\endgroup\@href {#1}{\urlprefix }}%
\providecommand \urlprefix  [0]{URL }%
\providecommand \Eprint [0]{\href }%
\providecommand \doibase [0]{http://dx.doi.org/}%
\providecommand \selectlanguage [0]{\@gobble}%
\providecommand \bibinfo  [0]{\@secondoftwo}%
\providecommand \bibfield  [0]{\@secondoftwo}%
\providecommand \translation [1]{[#1]}%
\providecommand \BibitemOpen [0]{}%
\providecommand \bibitemStop [0]{}%
\providecommand \bibitemNoStop [0]{.\EOS\space}%
\providecommand \EOS [0]{\spacefactor3000\relax}%
\providecommand \BibitemShut  [1]{\csname bibitem#1\endcsname}%
\let\auto@bib@innerbib\@empty
\bibitem [{\citenamefont {Senthil}\ \emph
  {et~al.}(2004{\natexlab{a}})\citenamefont {Senthil}, \citenamefont
  {Vishwanath}, \citenamefont {Balents}, \citenamefont {Sachdev},\ and\
  \citenamefont {Fisher}}]{dqcp}%
  \BibitemOpen
  \bibfield  {author} {\bibinfo {author} {\bibfnamefont {T.}~\bibnamefont
  {Senthil}}, \bibinfo {author} {\bibfnamefont {A.}~\bibnamefont {Vishwanath}},
  \bibinfo {author} {\bibfnamefont {L.}~\bibnamefont {Balents}}, \bibinfo
  {author} {\bibfnamefont {S.}~\bibnamefont {Sachdev}}, \ and\ \bibinfo
  {author} {\bibfnamefont {M.~P.~A.}\ \bibnamefont {Fisher}},\ }\href {\doibase
  10.1126/science.1091806} {\bibfield  {journal} {\bibinfo  {journal}
  {Science}\ }\textbf {\bibinfo {volume} {303}},\ \bibinfo {pages} {1490}
  (\bibinfo {year} {2004}{\natexlab{a}})},\ \Eprint
  {http://arxiv.org/abs/https://science.sciencemag.org/content/303/5663/1490.full.pdf}
  {https://science.sciencemag.org/content/303/5663/1490.full.pdf} \BibitemShut
  {NoStop}%
\bibitem [{\citenamefont {Senthil}\ \emph
  {et~al.}(2004{\natexlab{b}})\citenamefont {Senthil}, \citenamefont {Balents},
  \citenamefont {Sachdev}, \citenamefont {Vishwanath},\ and\ \citenamefont
  {Fisher}}]{dqcp2}%
  \BibitemOpen
  \bibfield  {author} {\bibinfo {author} {\bibfnamefont {T.}~\bibnamefont
  {Senthil}}, \bibinfo {author} {\bibfnamefont {L.}~\bibnamefont {Balents}},
  \bibinfo {author} {\bibfnamefont {S.}~\bibnamefont {Sachdev}}, \bibinfo
  {author} {\bibfnamefont {A.}~\bibnamefont {Vishwanath}}, \ and\ \bibinfo
  {author} {\bibfnamefont {M.~P.~A.}\ \bibnamefont {Fisher}},\ }\href {\doibase
  10.1103/PhysRevB.70.144407} {\bibfield  {journal} {\bibinfo  {journal} {Phys.
  Rev. B}\ }\textbf {\bibinfo {volume} {70}},\ \bibinfo {pages} {144407}
  (\bibinfo {year} {2004}{\natexlab{b}})}\BibitemShut {NoStop}%
\bibitem [{\citenamefont {Senthil}\ \emph {et~al.}(2005)\citenamefont
  {Senthil}, \citenamefont {Balents}, \citenamefont {Sachdev}, \citenamefont
  {Vishwanath},\ and\ \citenamefont {P.~A.~Fisher}}]{defined}%
  \BibitemOpen
  \bibfield  {author} {\bibinfo {author} {\bibfnamefont {T.}~\bibnamefont
  {Senthil}}, \bibinfo {author} {\bibfnamefont {L.}~\bibnamefont {Balents}},
  \bibinfo {author} {\bibfnamefont {S.}~\bibnamefont {Sachdev}}, \bibinfo
  {author} {\bibfnamefont {A.}~\bibnamefont {Vishwanath}}, \ and\ \bibinfo
  {author} {\bibfnamefont {M.}~\bibnamefont {P.~A.~Fisher}},\ }\href {\doibase
  10.1143/JPSJS.74S.1} {\bibfield  {journal} {\bibinfo  {journal} {Journal of
  the Physical Society of Japan}\ }\textbf {\bibinfo {volume} {74}},\ \bibinfo
  {pages} {1} (\bibinfo {year} {2005})},\ \Eprint
  {http://arxiv.org/abs/https://doi.org/10.1143/JPSJS.74S.1}
  {https://doi.org/10.1143/JPSJS.74S.1} \BibitemShut {NoStop}%
\bibitem [{\citenamefont {Levin}\ and\ \citenamefont
  {Senthil}(2004)}]{levinsenthil}%
  \BibitemOpen
  \bibfield  {author} {\bibinfo {author} {\bibfnamefont {M.}~\bibnamefont
  {Levin}}\ and\ \bibinfo {author} {\bibfnamefont {T.}~\bibnamefont
  {Senthil}},\ }\href {\doibase 10.1103/PhysRevB.70.220403} {\bibfield
  {journal} {\bibinfo  {journal} {Phys. Rev. B}\ }\textbf {\bibinfo {volume}
  {70}},\ \bibinfo {pages} {220403} (\bibinfo {year} {2004})}\BibitemShut
  {NoStop}%
\bibitem [{\citenamefont {Motrunich}\ and\ \citenamefont
  {Vishwanath}(2004)}]{hedgehog}%
  \BibitemOpen
  \bibfield  {author} {\bibinfo {author} {\bibfnamefont {O.~I.}\ \bibnamefont
  {Motrunich}}\ and\ \bibinfo {author} {\bibfnamefont {A.}~\bibnamefont
  {Vishwanath}},\ }\href {\doibase 10.1103/PhysRevB.70.075104} {\bibfield
  {journal} {\bibinfo  {journal} {Phys. Rev. B}\ }\textbf {\bibinfo {volume}
  {70}},\ \bibinfo {pages} {075104} (\bibinfo {year} {2004})}\BibitemShut
  {NoStop}%
\bibitem [{\citenamefont {Sandvik}(2007)}]{sandvik}%
  \BibitemOpen
  \bibfield  {author} {\bibinfo {author} {\bibfnamefont {A.~W.}\ \bibnamefont
  {Sandvik}},\ }\href {\doibase 10.1103/PhysRevLett.98.227202} {\bibfield
  {journal} {\bibinfo  {journal} {Phys. Rev. Lett.}\ }\textbf {\bibinfo
  {volume} {98}},\ \bibinfo {pages} {227202} (\bibinfo {year}
  {2007})}\BibitemShut {NoStop}%
\bibitem [{\citenamefont {Melko}\ and\ \citenamefont {Kaul}(2008)}]{melko}%
  \BibitemOpen
  \bibfield  {author} {\bibinfo {author} {\bibfnamefont {R.~G.}\ \bibnamefont
  {Melko}}\ and\ \bibinfo {author} {\bibfnamefont {R.~K.}\ \bibnamefont
  {Kaul}},\ }\href {\doibase 10.1103/PhysRevLett.100.017203} {\bibfield
  {journal} {\bibinfo  {journal} {Phys. Rev. Lett.}\ }\textbf {\bibinfo
  {volume} {100}},\ \bibinfo {pages} {017203} (\bibinfo {year}
  {2008})}\BibitemShut {NoStop}%
\bibitem [{\citenamefont {Lou}\ \emph {et~al.}(2009)\citenamefont {Lou},
  \citenamefont {Sandvik},\ and\ \citenamefont {Kawashima}}]{sandvik2}%
  \BibitemOpen
  \bibfield  {author} {\bibinfo {author} {\bibfnamefont {J.}~\bibnamefont
  {Lou}}, \bibinfo {author} {\bibfnamefont {A.~W.}\ \bibnamefont {Sandvik}}, \
  and\ \bibinfo {author} {\bibfnamefont {N.}~\bibnamefont {Kawashima}},\ }\href
  {\doibase 10.1103/PhysRevB.80.180414} {\bibfield  {journal} {\bibinfo
  {journal} {Phys. Rev. B}\ }\textbf {\bibinfo {volume} {80}},\ \bibinfo
  {pages} {180414} (\bibinfo {year} {2009})}\BibitemShut {NoStop}%
\bibitem [{\citenamefont {Banerjee}\ \emph {et~al.}(2010)\citenamefont
  {Banerjee}, \citenamefont {Damle},\ and\ \citenamefont {Alet}}]{banerjee}%
  \BibitemOpen
  \bibfield  {author} {\bibinfo {author} {\bibfnamefont {A.}~\bibnamefont
  {Banerjee}}, \bibinfo {author} {\bibfnamefont {K.}~\bibnamefont {Damle}}, \
  and\ \bibinfo {author} {\bibfnamefont {F.}~\bibnamefont {Alet}},\ }\href
  {\doibase 10.1103/PhysRevB.82.155139} {\bibfield  {journal} {\bibinfo
  {journal} {Phys. Rev. B}\ }\textbf {\bibinfo {volume} {82}},\ \bibinfo
  {pages} {155139} (\bibinfo {year} {2010})}\BibitemShut {NoStop}%
\bibitem [{\citenamefont {Sandvik}(2010)}]{sandvik3}%
  \BibitemOpen
  \bibfield  {author} {\bibinfo {author} {\bibfnamefont {A.~W.}\ \bibnamefont
  {Sandvik}},\ }\href {\doibase 10.1103/PhysRevLett.104.177201} {\bibfield
  {journal} {\bibinfo  {journal} {Phys. Rev. Lett.}\ }\textbf {\bibinfo
  {volume} {104}},\ \bibinfo {pages} {177201} (\bibinfo {year}
  {2010})}\BibitemShut {NoStop}%
\bibitem [{\citenamefont {Harada}\ \emph {et~al.}(2013)\citenamefont {Harada},
  \citenamefont {Suzuki}, \citenamefont {Okubo}, \citenamefont {Matsuo},
  \citenamefont {Lou}, \citenamefont {Watanabe}, \citenamefont {Todo},\ and\
  \citenamefont {Kawashima}}]{harada}%
  \BibitemOpen
  \bibfield  {author} {\bibinfo {author} {\bibfnamefont {K.}~\bibnamefont
  {Harada}}, \bibinfo {author} {\bibfnamefont {T.}~\bibnamefont {Suzuki}},
  \bibinfo {author} {\bibfnamefont {T.}~\bibnamefont {Okubo}}, \bibinfo
  {author} {\bibfnamefont {H.}~\bibnamefont {Matsuo}}, \bibinfo {author}
  {\bibfnamefont {J.}~\bibnamefont {Lou}}, \bibinfo {author} {\bibfnamefont
  {H.}~\bibnamefont {Watanabe}}, \bibinfo {author} {\bibfnamefont
  {S.}~\bibnamefont {Todo}}, \ and\ \bibinfo {author} {\bibfnamefont
  {N.}~\bibnamefont {Kawashima}},\ }\href {\doibase 10.1103/PhysRevB.88.220408}
  {\bibfield  {journal} {\bibinfo  {journal} {Phys. Rev. B}\ }\textbf {\bibinfo
  {volume} {88}},\ \bibinfo {pages} {220408} (\bibinfo {year}
  {2013})}\BibitemShut {NoStop}%
\bibitem [{\citenamefont {Pujari}\ \emph {et~al.}(2013)\citenamefont {Pujari},
  \citenamefont {Damle},\ and\ \citenamefont {Alet}}]{pujari}%
  \BibitemOpen
  \bibfield  {author} {\bibinfo {author} {\bibfnamefont {S.}~\bibnamefont
  {Pujari}}, \bibinfo {author} {\bibfnamefont {K.}~\bibnamefont {Damle}}, \
  and\ \bibinfo {author} {\bibfnamefont {F.}~\bibnamefont {Alet}},\ }\href
  {\doibase 10.1103/PhysRevLett.111.087203} {\bibfield  {journal} {\bibinfo
  {journal} {Phys. Rev. Lett.}\ }\textbf {\bibinfo {volume} {111}},\ \bibinfo
  {pages} {087203} (\bibinfo {year} {2013})}\BibitemShut {NoStop}%
\bibitem [{\citenamefont {Kaul}\ and\ \citenamefont
  {Sandvik}(2012)}]{sandvik4}%
  \BibitemOpen
  \bibfield  {author} {\bibinfo {author} {\bibfnamefont {R.~K.}\ \bibnamefont
  {Kaul}}\ and\ \bibinfo {author} {\bibfnamefont {A.~W.}\ \bibnamefont
  {Sandvik}},\ }\href {\doibase 10.1103/PhysRevLett.108.137201} {\bibfield
  {journal} {\bibinfo  {journal} {Phys. Rev. Lett.}\ }\textbf {\bibinfo
  {volume} {108}},\ \bibinfo {pages} {137201} (\bibinfo {year}
  {2012})}\BibitemShut {NoStop}%
\bibitem [{\citenamefont {Bartosch}(2013)}]{bartosch}%
  \BibitemOpen
  \bibfield  {author} {\bibinfo {author} {\bibfnamefont {L.}~\bibnamefont
  {Bartosch}},\ }\href {\doibase 10.1103/PhysRevB.88.195140} {\bibfield
  {journal} {\bibinfo  {journal} {Phys. Rev. B}\ }\textbf {\bibinfo {volume}
  {88}},\ \bibinfo {pages} {195140} (\bibinfo {year} {2013})}\BibitemShut
  {NoStop}%
\bibitem [{\citenamefont {Banerjee}\ \emph {et~al.}(2011)\citenamefont
  {Banerjee}, \citenamefont {Damle},\ and\ \citenamefont {Alet}}]{banerjee2}%
  \BibitemOpen
  \bibfield  {author} {\bibinfo {author} {\bibfnamefont {A.}~\bibnamefont
  {Banerjee}}, \bibinfo {author} {\bibfnamefont {K.}~\bibnamefont {Damle}}, \
  and\ \bibinfo {author} {\bibfnamefont {F.}~\bibnamefont {Alet}},\ }\href
  {\doibase 10.1103/PhysRevB.83.235111} {\bibfield  {journal} {\bibinfo
  {journal} {Phys. Rev. B}\ }\textbf {\bibinfo {volume} {83}},\ \bibinfo
  {pages} {235111} (\bibinfo {year} {2011})}\BibitemShut {NoStop}%
\bibitem [{\citenamefont {Nahum}\ \emph {et~al.}(2015)\citenamefont {Nahum},
  \citenamefont {Chalker}, \citenamefont {Serna}, \citenamefont {Ortu\~no},\
  and\ \citenamefont {Somoza}}]{nahumdeconfined}%
  \BibitemOpen
  \bibfield  {author} {\bibinfo {author} {\bibfnamefont {A.}~\bibnamefont
  {Nahum}}, \bibinfo {author} {\bibfnamefont {J.~T.}\ \bibnamefont {Chalker}},
  \bibinfo {author} {\bibfnamefont {P.}~\bibnamefont {Serna}}, \bibinfo
  {author} {\bibfnamefont {M.}~\bibnamefont {Ortu\~no}}, \ and\ \bibinfo
  {author} {\bibfnamefont {A.~M.}\ \bibnamefont {Somoza}},\ }\href {\doibase
  10.1103/PhysRevX.5.041048} {\bibfield  {journal} {\bibinfo  {journal} {Phys.
  Rev. X}\ }\textbf {\bibinfo {volume} {5}},\ \bibinfo {pages} {041048}
  (\bibinfo {year} {2015})}\BibitemShut {NoStop}%
\bibitem [{\citenamefont {Wang}\ \emph {et~al.}(2017)\citenamefont {Wang},
  \citenamefont {Nahum}, \citenamefont {Metlitski}, \citenamefont {Xu},\ and\
  \citenamefont {Senthil}}]{dualdeconfined}%
  \BibitemOpen
  \bibfield  {author} {\bibinfo {author} {\bibfnamefont {C.}~\bibnamefont
  {Wang}}, \bibinfo {author} {\bibfnamefont {A.}~\bibnamefont {Nahum}},
  \bibinfo {author} {\bibfnamefont {M.~A.}\ \bibnamefont {Metlitski}}, \bibinfo
  {author} {\bibfnamefont {C.}~\bibnamefont {Xu}}, \ and\ \bibinfo {author}
  {\bibfnamefont {T.}~\bibnamefont {Senthil}},\ }\href {\doibase
  10.1103/PhysRevX.7.031051} {\bibfield  {journal} {\bibinfo  {journal} {Phys.
  Rev. X}\ }\textbf {\bibinfo {volume} {7}},\ \bibinfo {pages} {031051}
  (\bibinfo {year} {2017})}\BibitemShut {NoStop}%
\bibitem [{\citenamefont {Vishwanath}\ \emph {et~al.}(2004)\citenamefont
  {Vishwanath}, \citenamefont {Balents},\ and\ \citenamefont
  {Senthil}}]{vbsdeconfined}%
  \BibitemOpen
  \bibfield  {author} {\bibinfo {author} {\bibfnamefont {A.}~\bibnamefont
  {Vishwanath}}, \bibinfo {author} {\bibfnamefont {L.}~\bibnamefont {Balents}},
  \ and\ \bibinfo {author} {\bibfnamefont {T.}~\bibnamefont {Senthil}},\ }\href
  {\doibase 10.1103/PhysRevB.69.224416} {\bibfield  {journal} {\bibinfo
  {journal} {Phys. Rev. B}\ }\textbf {\bibinfo {volume} {69}},\ \bibinfo
  {pages} {224416} (\bibinfo {year} {2004})}\BibitemShut {NoStop}%
\bibitem [{\citenamefont {Pankov}\ \emph
  {et~al.}(2007{\natexlab{a}})\citenamefont {Pankov}, \citenamefont
  {Moessner},\ and\ \citenamefont {Sondhi}}]{rsvp}%
  \BibitemOpen
  \bibfield  {author} {\bibinfo {author} {\bibfnamefont {S.}~\bibnamefont
  {Pankov}}, \bibinfo {author} {\bibfnamefont {R.}~\bibnamefont {Moessner}}, \
  and\ \bibinfo {author} {\bibfnamefont {S.~L.}\ \bibnamefont {Sondhi}},\
  }\href {\doibase 10.1103/PhysRevB.76.104436} {\bibfield  {journal} {\bibinfo
  {journal} {Phys. Rev. B}\ }\textbf {\bibinfo {volume} {76}},\ \bibinfo
  {pages} {104436} (\bibinfo {year} {2007}{\natexlab{a}})}\BibitemShut
  {NoStop}%
\bibitem [{\citenamefont {Gong}\ \emph {et~al.}(2014)\citenamefont {Gong},
  \citenamefont {Zhu}, \citenamefont {Sheng}, \citenamefont {Motrunich},\ and\
  \citenamefont {Fisher}}]{sheng}%
  \BibitemOpen
  \bibfield  {author} {\bibinfo {author} {\bibfnamefont {S.-S.}\ \bibnamefont
  {Gong}}, \bibinfo {author} {\bibfnamefont {W.}~\bibnamefont {Zhu}}, \bibinfo
  {author} {\bibfnamefont {D.~N.}\ \bibnamefont {Sheng}}, \bibinfo {author}
  {\bibfnamefont {O.~I.}\ \bibnamefont {Motrunich}}, \ and\ \bibinfo {author}
  {\bibfnamefont {M.~P.~A.}\ \bibnamefont {Fisher}},\ }\href {\doibase
  10.1103/PhysRevLett.113.027201} {\bibfield  {journal} {\bibinfo  {journal}
  {Phys. Rev. Lett.}\ }\textbf {\bibinfo {volume} {113}},\ \bibinfo {pages}
  {027201} (\bibinfo {year} {2014})}\BibitemShut {NoStop}%
\bibitem [{\citenamefont {Xu}\ and\ \citenamefont
  {Wu}(2008)}]{xu2008resonating}%
  \BibitemOpen
  \bibfield  {author} {\bibinfo {author} {\bibfnamefont {C.}~\bibnamefont
  {Xu}}\ and\ \bibinfo {author} {\bibfnamefont {C.}~\bibnamefont {Wu}},\
  }\href@noop {} {\bibfield  {journal} {\bibinfo  {journal} {Physical Review
  B}\ }\textbf {\bibinfo {volume} {77}},\ \bibinfo {pages} {134449} (\bibinfo
  {year} {2008})}\BibitemShut {NoStop}%
\bibitem [{\citenamefont {Zhu}\ \emph {et~al.}(2013)\citenamefont {Zhu},
  \citenamefont {Huse},\ and\ \citenamefont {White}}]{zhuwhite}%
  \BibitemOpen
  \bibfield  {author} {\bibinfo {author} {\bibfnamefont {Z.}~\bibnamefont
  {Zhu}}, \bibinfo {author} {\bibfnamefont {D.~A.}\ \bibnamefont {Huse}}, \
  and\ \bibinfo {author} {\bibfnamefont {S.~R.}\ \bibnamefont {White}},\ }\href
  {\doibase 10.1103/PhysRevLett.110.127205} {\bibfield  {journal} {\bibinfo
  {journal} {Phys. Rev. Lett.}\ }\textbf {\bibinfo {volume} {110}},\ \bibinfo
  {pages} {127205} (\bibinfo {year} {2013})}\BibitemShut {NoStop}%
\bibitem [{\citenamefont {Moessner}\ \emph {et~al.}(2001)\citenamefont
  {Moessner}, \citenamefont {Sondhi},\ and\ \citenamefont
  {Chandra}}]{hexdimer}%
  \BibitemOpen
  \bibfield  {author} {\bibinfo {author} {\bibfnamefont {R.}~\bibnamefont
  {Moessner}}, \bibinfo {author} {\bibfnamefont {S.~L.}\ \bibnamefont
  {Sondhi}}, \ and\ \bibinfo {author} {\bibfnamefont {P.}~\bibnamefont
  {Chandra}},\ }\href {\doibase 10.1103/PhysRevB.64.144416} {\bibfield
  {journal} {\bibinfo  {journal} {Phys. Rev. B}\ }\textbf {\bibinfo {volume}
  {64}},\ \bibinfo {pages} {144416} (\bibinfo {year} {2001})}\BibitemShut
  {NoStop}%
\bibitem [{\citenamefont {Zhao}\ \emph {et~al.}(2012)\citenamefont {Zhao},
  \citenamefont {Xu}, \citenamefont {Chen}, \citenamefont {Wei}, \citenamefont
  {Qin}, \citenamefont {Zhang},\ and\ \citenamefont {Xiang}}]{zhaoplaq}%
  \BibitemOpen
  \bibfield  {author} {\bibinfo {author} {\bibfnamefont {H.~H.}\ \bibnamefont
  {Zhao}}, \bibinfo {author} {\bibfnamefont {C.}~\bibnamefont {Xu}}, \bibinfo
  {author} {\bibfnamefont {Q.~N.}\ \bibnamefont {Chen}}, \bibinfo {author}
  {\bibfnamefont {Z.~C.}\ \bibnamefont {Wei}}, \bibinfo {author} {\bibfnamefont
  {M.~P.}\ \bibnamefont {Qin}}, \bibinfo {author} {\bibfnamefont {G.~M.}\
  \bibnamefont {Zhang}}, \ and\ \bibinfo {author} {\bibfnamefont
  {T.}~\bibnamefont {Xiang}},\ }\href {\doibase 10.1103/PhysRevB.85.134416}
  {\bibfield  {journal} {\bibinfo  {journal} {Phys. Rev. B}\ }\textbf {\bibinfo
  {volume} {85}},\ \bibinfo {pages} {134416} (\bibinfo {year}
  {2012})}\BibitemShut {NoStop}%
\bibitem [{\citenamefont {Albuquerque}\ \emph {et~al.}(2011)\citenamefont
  {Albuquerque}, \citenamefont {Schwandt}, \citenamefont {Het\'enyi},
  \citenamefont {Capponi}, \citenamefont {Mambrini},\ and\ \citenamefont
  {L\"auchli}}]{albuq}%
  \BibitemOpen
  \bibfield  {author} {\bibinfo {author} {\bibfnamefont {A.~F.}\ \bibnamefont
  {Albuquerque}}, \bibinfo {author} {\bibfnamefont {D.}~\bibnamefont
  {Schwandt}}, \bibinfo {author} {\bibfnamefont {B.}~\bibnamefont {Het\'enyi}},
  \bibinfo {author} {\bibfnamefont {S.}~\bibnamefont {Capponi}}, \bibinfo
  {author} {\bibfnamefont {M.}~\bibnamefont {Mambrini}}, \ and\ \bibinfo
  {author} {\bibfnamefont {A.~M.}\ \bibnamefont {L\"auchli}},\ }\href {\doibase
  10.1103/PhysRevB.84.024406} {\bibfield  {journal} {\bibinfo  {journal} {Phys.
  Rev. B}\ }\textbf {\bibinfo {volume} {84}},\ \bibinfo {pages} {024406}
  (\bibinfo {year} {2011})}\BibitemShut {NoStop}%
\bibitem [{\citenamefont {Zayed}\ \emph {et~al.}(2017)\citenamefont {Zayed},
  \citenamefont {R{\"u}egg}, \citenamefont {L{\"a}uchli}, \citenamefont
  {Panagopoulos}, \citenamefont {Saxena}, \citenamefont {Ellerby},
  \citenamefont {McMorrow}, \citenamefont {Str{\"a}ssle}, \citenamefont
  {Klotz}, \citenamefont {Hamel} \emph {et~al.}}]{zayed20174}%
  \BibitemOpen
  \bibfield  {author} {\bibinfo {author} {\bibfnamefont {M.}~\bibnamefont
  {Zayed}}, \bibinfo {author} {\bibfnamefont {C.}~\bibnamefont {R{\"u}egg}},
  \bibinfo {author} {\bibfnamefont {A.}~\bibnamefont {L{\"a}uchli}}, \bibinfo
  {author} {\bibfnamefont {C.}~\bibnamefont {Panagopoulos}}, \bibinfo {author}
  {\bibfnamefont {S.}~\bibnamefont {Saxena}}, \bibinfo {author} {\bibfnamefont
  {M.}~\bibnamefont {Ellerby}}, \bibinfo {author} {\bibfnamefont
  {D.}~\bibnamefont {McMorrow}}, \bibinfo {author} {\bibfnamefont
  {T.}~\bibnamefont {Str{\"a}ssle}}, \bibinfo {author} {\bibfnamefont
  {S.}~\bibnamefont {Klotz}}, \bibinfo {author} {\bibfnamefont
  {G.}~\bibnamefont {Hamel}},  \emph {et~al.},\ }\href@noop {} {\bibfield
  {journal} {\bibinfo  {journal} {Nature Physics}\ }\textbf {\bibinfo {volume}
  {13}},\ \bibinfo {pages} {962} (\bibinfo {year} {2017})}\BibitemShut
  {NoStop}%
\bibitem [{\citenamefont {Zhao}\ \emph {et~al.}(2018)\citenamefont {Zhao},
  \citenamefont {Weinberg},\ and\ \citenamefont {Sandvik}}]{zhao2018symmetry}%
  \BibitemOpen
  \bibfield  {author} {\bibinfo {author} {\bibfnamefont {B.}~\bibnamefont
  {Zhao}}, \bibinfo {author} {\bibfnamefont {P.}~\bibnamefont {Weinberg}}, \
  and\ \bibinfo {author} {\bibfnamefont {A.~W.}\ \bibnamefont {Sandvik}},\
  }\href@noop {} {\bibfield  {journal} {\bibinfo  {journal} {arXiv preprint
  arXiv:1804.07115}\ } (\bibinfo {year} {2018})}\BibitemShut {NoStop}%
\bibitem [{\citenamefont {Nandkishore}\ and\ \citenamefont
  {Hermele}(2019)}]{review}%
  \BibitemOpen
  \bibfield  {author} {\bibinfo {author} {\bibfnamefont {R.~M.}\ \bibnamefont
  {Nandkishore}}\ and\ \bibinfo {author} {\bibfnamefont {M.}~\bibnamefont
  {Hermele}},\ }\href {\doibase 10.1146/annurev-conmatphys-031218-013604}
  {\bibfield  {journal} {\bibinfo  {journal} {Annual Review of Condensed Matter
  Physics}\ }\textbf {\bibinfo {volume} {10}},\ \bibinfo {pages} {295}
  (\bibinfo {year} {2019})},\ \Eprint
  {http://arxiv.org/abs/https://doi.org/10.1146/annurev-conmatphys-031218-013604}
  {https://doi.org/10.1146/annurev-conmatphys-031218-013604} \BibitemShut
  {NoStop}%
\bibitem [{\citenamefont {Vijay}\ \emph {et~al.}(2015)\citenamefont {Vijay},
  \citenamefont {Haah},\ and\ \citenamefont {Fu}}]{Vijay2015-jj}%
  \BibitemOpen
  \bibfield  {author} {\bibinfo {author} {\bibfnamefont {S.}~\bibnamefont
  {Vijay}}, \bibinfo {author} {\bibfnamefont {J.}~\bibnamefont {Haah}}, \ and\
  \bibinfo {author} {\bibfnamefont {L.}~\bibnamefont {Fu}},\ }\href@noop {}
  {\bibfield  {journal} {\bibinfo  {journal} {Phys. Rev. B}\ }\textbf {\bibinfo
  {volume} {92}},\ \bibinfo {pages} {235136} (\bibinfo {year}
  {2015})}\BibitemShut {NoStop}%
\bibitem [{\citenamefont {Vijay}\ \emph {et~al.}(2016)\citenamefont {Vijay},
  \citenamefont {Haah},\ and\ \citenamefont {Fu}}]{Vijay2016-dr}%
  \BibitemOpen
  \bibfield  {author} {\bibinfo {author} {\bibfnamefont {S.}~\bibnamefont
  {Vijay}}, \bibinfo {author} {\bibfnamefont {J.}~\bibnamefont {Haah}}, \ and\
  \bibinfo {author} {\bibfnamefont {L.}~\bibnamefont {Fu}},\ }\href@noop {}
  {\bibfield  {journal} {\bibinfo  {journal} {Phys. Rev. B}\ }\textbf {\bibinfo
  {volume} {94}},\ \bibinfo {pages} {235157} (\bibinfo {year}
  {2016})}\BibitemShut {NoStop}%
\bibitem [{\citenamefont {Chamon}(2005)}]{Chamon2005-fc}%
  \BibitemOpen
  \bibfield  {author} {\bibinfo {author} {\bibfnamefont {C.}~\bibnamefont
  {Chamon}},\ }\href@noop {} {\bibfield  {journal} {\bibinfo  {journal} {Phys.
  Rev. Lett.}\ }\textbf {\bibinfo {volume} {94}},\ \bibinfo {pages} {040402}
  (\bibinfo {year} {2005})}\BibitemShut {NoStop}%
\bibitem [{\citenamefont {Haah}(2011)}]{Haah2011-ny}%
  \BibitemOpen
  \bibfield  {author} {\bibinfo {author} {\bibfnamefont {J.}~\bibnamefont
  {Haah}},\ }\href@noop {} {\bibfield  {journal} {\bibinfo  {journal} {Phys.
  Rev. A}\ }\textbf {\bibinfo {volume} {83}},\ \bibinfo {pages} {042330}
  (\bibinfo {year} {2011})}\BibitemShut {NoStop}%
\bibitem [{\citenamefont {Yoshida}(2013)}]{yoshida2013exotic}%
  \BibitemOpen
  \bibfield  {author} {\bibinfo {author} {\bibfnamefont {B.}~\bibnamefont
  {Yoshida}},\ }\href@noop {} {\bibfield  {journal} {\bibinfo  {journal} {Phys.
  Rev. B}\ }\textbf {\bibinfo {volume} {88}},\ \bibinfo {pages} {125122}
  (\bibinfo {year} {2013})}\BibitemShut {NoStop}%
\bibitem [{\citenamefont {Pretko}\ and\ \citenamefont
  {Radzihovsky}(2018{\natexlab{a}})}]{elasticity}%
  \BibitemOpen
  \bibfield  {author} {\bibinfo {author} {\bibfnamefont {M.}~\bibnamefont
  {Pretko}}\ and\ \bibinfo {author} {\bibfnamefont {L.}~\bibnamefont
  {Radzihovsky}},\ }\href@noop {} {\bibfield  {journal} {\bibinfo  {journal}
  {Phys. Rev. Lett.}\ }\textbf {\bibinfo {volume} {120}},\ \bibinfo {pages}
  {195301} (\bibinfo {year} {2018}{\natexlab{a}})}\BibitemShut {NoStop}%
\bibitem [{\citenamefont {Pai}\ and\ \citenamefont
  {Pretko}(2018)}]{pai2018fractonic}%
  \BibitemOpen
  \bibfield  {author} {\bibinfo {author} {\bibfnamefont {S.}~\bibnamefont
  {Pai}}\ and\ \bibinfo {author} {\bibfnamefont {M.}~\bibnamefont {Pretko}},\
  }\href {\doibase 10.1103/PhysRevB.97.235102} {\bibfield  {journal} {\bibinfo
  {journal} {Phys. Rev. B}\ }\textbf {\bibinfo {volume} {97}},\ \bibinfo
  {pages} {235102} (\bibinfo {year} {2018})}\BibitemShut {NoStop}%
\bibitem [{\citenamefont {Gromov}(2019)}]{gromov2017fractional}%
  \BibitemOpen
  \bibfield  {author} {\bibinfo {author} {\bibfnamefont {A.}~\bibnamefont
  {Gromov}},\ }\href {\doibase 10.1103/PhysRevLett.122.076403} {\bibfield
  {journal} {\bibinfo  {journal} {Phys. Rev. Lett.}\ }\textbf {\bibinfo
  {volume} {122}},\ \bibinfo {pages} {076403} (\bibinfo {year}
  {2019})}\BibitemShut {NoStop}%
\bibitem [{\citenamefont {Pretko}\ and\ \citenamefont
  {Radzihovsky}(2018{\natexlab{b}})}]{supersolid}%
  \BibitemOpen
  \bibfield  {author} {\bibinfo {author} {\bibfnamefont {M.}~\bibnamefont
  {Pretko}}\ and\ \bibinfo {author} {\bibfnamefont {L.}~\bibnamefont
  {Radzihovsky}},\ }\href {\doibase 10.1103/PhysRevLett.121.235301} {\bibfield
  {journal} {\bibinfo  {journal} {Phys. Rev. Lett.}\ }\textbf {\bibinfo
  {volume} {121}},\ \bibinfo {pages} {235301} (\bibinfo {year}
  {2018}{\natexlab{b}})}\BibitemShut {NoStop}%
\bibitem [{\citenamefont {{Kumar}}\ and\ \citenamefont
  {{Potter}}(2018)}]{kumarpotter}%
  \BibitemOpen
  \bibfield  {author} {\bibinfo {author} {\bibfnamefont {A.}~\bibnamefont
  {{Kumar}}}\ and\ \bibinfo {author} {\bibfnamefont {A.~C.}\ \bibnamefont
  {{Potter}}},\ }\href@noop {} {\bibfield  {journal} {\bibinfo  {journal}
  {arXiv e-prints}\ ,\ \bibinfo {eid} {arXiv:1808.05621}} (\bibinfo {year}
  {2018})},\ \Eprint {http://arxiv.org/abs/1808.05621} {arXiv:1808.05621
  [cond-mat.str-el]} \BibitemShut {NoStop}%
\bibitem [{\citenamefont {{Sous}}\ and\ \citenamefont
  {{Pretko}}(2019)}]{polaron}%
  \BibitemOpen
  \bibfield  {author} {\bibinfo {author} {\bibfnamefont {J.}~\bibnamefont
  {{Sous}}}\ and\ \bibinfo {author} {\bibfnamefont {M.}~\bibnamefont
  {{Pretko}}},\ }\href@noop {} {\bibfield  {journal} {\bibinfo  {journal}
  {arXiv e-prints}\ ,\ \bibinfo {eid} {arXiv:1904.08424}} (\bibinfo {year}
  {2019})},\ \Eprint {http://arxiv.org/abs/1904.08424} {arXiv:1904.08424
  [cond-mat.str-el]} \BibitemShut {NoStop}%
\bibitem [{\citenamefont {Shirley}\ \emph {et~al.}(2018)\citenamefont
  {Shirley}, \citenamefont {Slagle}, \citenamefont {Wang},\ and\ \citenamefont
  {Chen}}]{shirley2017fracton}%
  \BibitemOpen
  \bibfield  {author} {\bibinfo {author} {\bibfnamefont {W.}~\bibnamefont
  {Shirley}}, \bibinfo {author} {\bibfnamefont {K.}~\bibnamefont {Slagle}},
  \bibinfo {author} {\bibfnamefont {Z.}~\bibnamefont {Wang}}, \ and\ \bibinfo
  {author} {\bibfnamefont {X.}~\bibnamefont {Chen}},\ }\href {\doibase
  10.1103/PhysRevX.8.031051} {\bibfield  {journal} {\bibinfo  {journal} {Phys.
  Rev. X}\ }\textbf {\bibinfo {volume} {8}},\ \bibinfo {pages} {031051}
  (\bibinfo {year} {2018})}\BibitemShut {NoStop}%
\bibitem [{\citenamefont {{Pai}}\ and\ \citenamefont
  {{Hermele}}(2019)}]{paifusion}%
  \BibitemOpen
  \bibfield  {author} {\bibinfo {author} {\bibfnamefont {S.}~\bibnamefont
  {{Pai}}}\ and\ \bibinfo {author} {\bibfnamefont {M.}~\bibnamefont
  {{Hermele}}},\ }\href@noop {} {\bibfield  {journal} {\bibinfo  {journal}
  {arXiv e-prints}\ ,\ \bibinfo {eid} {arXiv:1903.11625}} (\bibinfo {year}
  {2019})},\ \Eprint {http://arxiv.org/abs/1903.11625} {arXiv:1903.11625
  [cond-mat.str-el]} \BibitemShut {NoStop}%
\bibitem [{\citenamefont {Slagle}\ \emph {et~al.}(2018)\citenamefont {Slagle},
  \citenamefont {Aasen},\ and\ \citenamefont
  {Williamson}}]{slagle2018foliated}%
  \BibitemOpen
  \bibfield  {author} {\bibinfo {author} {\bibfnamefont {K.}~\bibnamefont
  {Slagle}}, \bibinfo {author} {\bibfnamefont {D.}~\bibnamefont {Aasen}}, \
  and\ \bibinfo {author} {\bibfnamefont {D.}~\bibnamefont {Williamson}},\
  }\href@noop {} {\bibfield  {journal} {\bibinfo  {journal} {arXiv preprint
  arXiv:1812.01613}\ } (\bibinfo {year} {2018})}\BibitemShut {NoStop}%
\bibitem [{\citenamefont {Shirley}\ \emph {et~al.}(2019)\citenamefont
  {Shirley}, \citenamefont {Slagle},\ and\ \citenamefont
  {Chen}}]{shirley2018foliated}%
  \BibitemOpen
  \bibfield  {author} {\bibinfo {author} {\bibfnamefont {W.}~\bibnamefont
  {Shirley}}, \bibinfo {author} {\bibfnamefont {K.}~\bibnamefont {Slagle}}, \
  and\ \bibinfo {author} {\bibfnamefont {X.}~\bibnamefont {Chen}},\ }\href
  {\doibase 10.21468/SciPostPhys.6.4.041} {\bibfield  {journal} {\bibinfo
  {journal} {SciPost Phys.}\ }\textbf {\bibinfo {volume} {6}},\ \bibinfo
  {pages} {41} (\bibinfo {year} {2019})}\BibitemShut {NoStop}%
\bibitem [{\citenamefont {You}\ \emph {et~al.}(2018)\citenamefont {You},
  \citenamefont {Devakul}, \citenamefont {Burnell},\ and\ \citenamefont
  {Sondhi}}]{you2018symmetric}%
  \BibitemOpen
  \bibfield  {author} {\bibinfo {author} {\bibfnamefont {Y.}~\bibnamefont
  {You}}, \bibinfo {author} {\bibfnamefont {T.}~\bibnamefont {Devakul}},
  \bibinfo {author} {\bibfnamefont {F.}~\bibnamefont {Burnell}}, \ and\
  \bibinfo {author} {\bibfnamefont {S.}~\bibnamefont {Sondhi}},\ }\href@noop {}
  {\bibfield  {journal} {\bibinfo  {journal} {arXiv preprint arXiv:1805.09800}\
  } (\bibinfo {year} {2018})}\BibitemShut {NoStop}%
\bibitem [{\citenamefont {Song}\ \emph {et~al.}(2019)\citenamefont {Song},
  \citenamefont {Prem}, \citenamefont {Huang},\ and\ \citenamefont
  {Martin-Delgado}}]{song2018twisted}%
  \BibitemOpen
  \bibfield  {author} {\bibinfo {author} {\bibfnamefont {H.}~\bibnamefont
  {Song}}, \bibinfo {author} {\bibfnamefont {A.}~\bibnamefont {Prem}}, \bibinfo
  {author} {\bibfnamefont {S.-J.}\ \bibnamefont {Huang}}, \ and\ \bibinfo
  {author} {\bibfnamefont {M.~A.}\ \bibnamefont {Martin-Delgado}},\ }\href
  {\doibase 10.1103/PhysRevB.99.155118} {\bibfield  {journal} {\bibinfo
  {journal} {Phys. Rev. B}\ }\textbf {\bibinfo {volume} {99}},\ \bibinfo
  {pages} {155118} (\bibinfo {year} {2019})}\BibitemShut {NoStop}%
\bibitem [{\citenamefont {Slagle}\ and\ \citenamefont
  {Kim}(2017)}]{Slagle2017-gk}%
  \BibitemOpen
  \bibfield  {author} {\bibinfo {author} {\bibfnamefont {K.}~\bibnamefont
  {Slagle}}\ and\ \bibinfo {author} {\bibfnamefont {Y.~B.}\ \bibnamefont
  {Kim}},\ }\href@noop {} {\bibfield  {journal} {\bibinfo  {journal} {Phys.
  Rev. B Condens. Matter}\ }\textbf {\bibinfo {volume} {96}},\ \bibinfo {pages}
  {195139} (\bibinfo {year} {2017})}\BibitemShut {NoStop}%
\bibitem [{\citenamefont {Ma}\ \emph {et~al.}(2017)\citenamefont {Ma},
  \citenamefont {Lake}, \citenamefont {Chen},\ and\ \citenamefont
  {Hermele}}]{Ma2017-qq}%
  \BibitemOpen
  \bibfield  {author} {\bibinfo {author} {\bibfnamefont {H.}~\bibnamefont
  {Ma}}, \bibinfo {author} {\bibfnamefont {E.}~\bibnamefont {Lake}}, \bibinfo
  {author} {\bibfnamefont {X.}~\bibnamefont {Chen}}, \ and\ \bibinfo {author}
  {\bibfnamefont {M.}~\bibnamefont {Hermele}},\ }\href {\doibase
  10.1103/PhysRevB.95.245126} {\bibfield  {journal} {\bibinfo  {journal} {Phys.
  Rev. B}\ }\textbf {\bibinfo {volume} {95}},\ \bibinfo {pages} {245126}
  (\bibinfo {year} {2017})}\BibitemShut {NoStop}%
\bibitem [{\citenamefont {Ma}\ \emph {et~al.}(2018{\natexlab{a}})\citenamefont
  {Ma}, \citenamefont {Schmitz}, \citenamefont {Parameswaran}, \citenamefont
  {Hermele},\ and\ \citenamefont {Nandkishore}}]{Ma2017-cb}%
  \BibitemOpen
  \bibfield  {author} {\bibinfo {author} {\bibfnamefont {H.}~\bibnamefont
  {Ma}}, \bibinfo {author} {\bibfnamefont {A.~T.}\ \bibnamefont {Schmitz}},
  \bibinfo {author} {\bibfnamefont {S.~A.}\ \bibnamefont {Parameswaran}},
  \bibinfo {author} {\bibfnamefont {M.}~\bibnamefont {Hermele}}, \ and\
  \bibinfo {author} {\bibfnamefont {R.~M.}\ \bibnamefont {Nandkishore}},\
  }\href {\doibase 10.1103/PhysRevB.97.125101} {\bibfield  {journal} {\bibinfo
  {journal} {Phys. Rev. B}\ }\textbf {\bibinfo {volume} {97}},\ \bibinfo
  {pages} {125101} (\bibinfo {year} {2018}{\natexlab{a}})}\BibitemShut
  {NoStop}%
\bibitem [{\citenamefont {Vijay}(2017)}]{Vijay2017-ey}%
  \BibitemOpen
  \bibfield  {author} {\bibinfo {author} {\bibfnamefont {S.}~\bibnamefont
  {Vijay}},\ }\href@noop {} {\  (\bibinfo {year} {2017})},\ \Eprint
  {http://arxiv.org/abs/1701.00762} {arXiv:1701.00762 [cond-mat.str-el]}
  \BibitemShut {NoStop}%
\bibitem [{\citenamefont {Petrova}\ and\ \citenamefont
  {Regnault}(2017)}]{Petrova2017-pe}%
  \BibitemOpen
  \bibfield  {author} {\bibinfo {author} {\bibfnamefont {O.}~\bibnamefont
  {Petrova}}\ and\ \bibinfo {author} {\bibfnamefont {N.}~\bibnamefont
  {Regnault}},\ }\href {\doibase 10.1103/PhysRevB.96.224429} {\bibfield
  {journal} {\bibinfo  {journal} {Phys. Rev. B}\ }\textbf {\bibinfo {volume}
  {96}},\ \bibinfo {pages} {224429} (\bibinfo {year} {2017})}\BibitemShut
  {NoStop}%
\bibitem [{\citenamefont {Schmitz}\ \emph {et~al.}(2018)\citenamefont
  {Schmitz}, \citenamefont {Ma}, \citenamefont {Nandkishore},\ and\
  \citenamefont {Parameswaran}}]{Schmitz2017-ky}%
  \BibitemOpen
  \bibfield  {author} {\bibinfo {author} {\bibfnamefont {A.~T.}\ \bibnamefont
  {Schmitz}}, \bibinfo {author} {\bibfnamefont {H.}~\bibnamefont {Ma}},
  \bibinfo {author} {\bibfnamefont {R.~M.}\ \bibnamefont {Nandkishore}}, \ and\
  \bibinfo {author} {\bibfnamefont {S.~A.}\ \bibnamefont {Parameswaran}},\
  }\href {\doibase 10.1103/PhysRevB.97.134426} {\bibfield  {journal} {\bibinfo
  {journal} {Phys. Rev. B}\ }\textbf {\bibinfo {volume} {97}},\ \bibinfo
  {pages} {134426} (\bibinfo {year} {2018})}\BibitemShut {NoStop}%
\bibitem [{\citenamefont {Schmitz}(2018)}]{schmitz2018gauge}%
  \BibitemOpen
  \bibfield  {author} {\bibinfo {author} {\bibfnamefont {A.~T.}\ \bibnamefont
  {Schmitz}},\ }\href@noop {} {\bibfield  {journal} {\bibinfo  {journal} {arXiv
  preprint arXiv:1809.10151}\ } (\bibinfo {year} {2018})}\BibitemShut {NoStop}%
\bibitem [{\citenamefont {Prem}\ \emph {et~al.}(2019)\citenamefont {Prem},
  \citenamefont {Huang}, \citenamefont {Song},\ and\ \citenamefont
  {Hermele}}]{huang2018cage}%
  \BibitemOpen
  \bibfield  {author} {\bibinfo {author} {\bibfnamefont {A.}~\bibnamefont
  {Prem}}, \bibinfo {author} {\bibfnamefont {S.-J.}\ \bibnamefont {Huang}},
  \bibinfo {author} {\bibfnamefont {H.}~\bibnamefont {Song}}, \ and\ \bibinfo
  {author} {\bibfnamefont {M.}~\bibnamefont {Hermele}},\ }\href {\doibase
  10.1103/PhysRevX.9.021010} {\bibfield  {journal} {\bibinfo  {journal} {Phys.
  Rev. X}\ }\textbf {\bibinfo {volume} {9}},\ \bibinfo {pages} {021010}
  (\bibinfo {year} {2019})}\BibitemShut {NoStop}%
\bibitem [{\citenamefont {{Yan}}\ \emph {et~al.}(2019)\citenamefont {{Yan}},
  \citenamefont {{Benton}}, \citenamefont {{Jaubert}},\ and\ \citenamefont
  {{Shannon}}}]{rank2ice}%
  \BibitemOpen
  \bibfield  {author} {\bibinfo {author} {\bibfnamefont {H.}~\bibnamefont
  {{Yan}}}, \bibinfo {author} {\bibfnamefont {O.}~\bibnamefont {{Benton}}},
  \bibinfo {author} {\bibfnamefont {L.~D.~C.}\ \bibnamefont {{Jaubert}}}, \
  and\ \bibinfo {author} {\bibfnamefont {N.}~\bibnamefont {{Shannon}}},\
  }\href@noop {} {\bibfield  {journal} {\bibinfo  {journal} {arXiv e-prints}\
  ,\ \bibinfo {eid} {arXiv:1902.10934}} (\bibinfo {year} {2019})},\ \Eprint
  {http://arxiv.org/abs/1902.10934} {arXiv:1902.10934 [cond-mat.str-el]}
  \BibitemShut {NoStop}%
\bibitem [{\citenamefont {{Dua}}\ \emph {et~al.}(2019)\citenamefont {{Dua}},
  \citenamefont {{Williamson}}, \citenamefont {{Haah}},\ and\ \citenamefont
  {{Cheng}}}]{compactify}%
  \BibitemOpen
  \bibfield  {author} {\bibinfo {author} {\bibfnamefont {A.}~\bibnamefont
  {{Dua}}}, \bibinfo {author} {\bibfnamefont {D.~J.}\ \bibnamefont
  {{Williamson}}}, \bibinfo {author} {\bibfnamefont {J.}~\bibnamefont
  {{Haah}}}, \ and\ \bibinfo {author} {\bibfnamefont {M.}~\bibnamefont
  {{Cheng}}},\ }\href@noop {} {\bibfield  {journal} {\bibinfo  {journal} {arXiv
  e-prints}\ ,\ \bibinfo {eid} {arXiv:1903.12246}} (\bibinfo {year} {2019})},\
  \Eprint {http://arxiv.org/abs/1903.12246} {arXiv:1903.12246
  [cond-mat.str-el]} \BibitemShut {NoStop}%
\bibitem [{\citenamefont {Bravyi}\ and\ \citenamefont
  {Haah}(2013)}]{bravyihaah}%
  \BibitemOpen
  \bibfield  {author} {\bibinfo {author} {\bibfnamefont {S.}~\bibnamefont
  {Bravyi}}\ and\ \bibinfo {author} {\bibfnamefont {J.}~\bibnamefont {Haah}},\
  }\href {\doibase 10.1103/PhysRevLett.111.200501} {\bibfield  {journal}
  {\bibinfo  {journal} {Phys. Rev. Lett.}\ }\textbf {\bibinfo {volume} {111}},\
  \bibinfo {pages} {200501} (\bibinfo {year} {2013})}\BibitemShut {NoStop}%
\bibitem [{\citenamefont {Terhal}(2015)}]{terhal}%
  \BibitemOpen
  \bibfield  {author} {\bibinfo {author} {\bibfnamefont {B.~M.}\ \bibnamefont
  {Terhal}},\ }\href {\doibase 10.1103/RevModPhys.87.307} {\bibfield  {journal}
  {\bibinfo  {journal} {Rev. Mod. Phys.}\ }\textbf {\bibinfo {volume} {87}},\
  \bibinfo {pages} {307} (\bibinfo {year} {2015})}\BibitemShut {NoStop}%
\bibitem [{\citenamefont {Pretko}(2017{\natexlab{a}})}]{gravity}%
  \BibitemOpen
  \bibfield  {author} {\bibinfo {author} {\bibfnamefont {M.}~\bibnamefont
  {Pretko}},\ }\href {\doibase 10.1103/PhysRevD.96.024051} {\bibfield
  {journal} {\bibinfo  {journal} {Phys. Rev. D}\ }\textbf {\bibinfo {volume}
  {96}},\ \bibinfo {pages} {024051} (\bibinfo {year}
  {2017}{\natexlab{a}})}\BibitemShut {NoStop}%
\bibitem [{\citenamefont {Yan}(2019)}]{holo}%
  \BibitemOpen
  \bibfield  {author} {\bibinfo {author} {\bibfnamefont {H.}~\bibnamefont
  {Yan}},\ }\href {\doibase 10.1103/PhysRevB.99.155126} {\bibfield  {journal}
  {\bibinfo  {journal} {Phys. Rev. B}\ }\textbf {\bibinfo {volume} {99}},\
  \bibinfo {pages} {155126} (\bibinfo {year} {2019})}\BibitemShut {NoStop}%
\bibitem [{\citenamefont {Prem}\ \emph {et~al.}(2017)\citenamefont {Prem},
  \citenamefont {Haah},\ and\ \citenamefont {Nandkishore}}]{Prem2017-ql}%
  \BibitemOpen
  \bibfield  {author} {\bibinfo {author} {\bibfnamefont {A.}~\bibnamefont
  {Prem}}, \bibinfo {author} {\bibfnamefont {J.}~\bibnamefont {Haah}}, \ and\
  \bibinfo {author} {\bibfnamefont {R.}~\bibnamefont {Nandkishore}},\
  }\href@noop {} {\bibfield  {journal} {\bibinfo  {journal} {Phys. Rev. B
  Condens. Matter}\ }\textbf {\bibinfo {volume} {95}},\ \bibinfo {pages}
  {155133} (\bibinfo {year} {2017})}\BibitemShut {NoStop}%
\bibitem [{\citenamefont {Pai}\ \emph {et~al.}(2019)\citenamefont {Pai},
  \citenamefont {Pretko},\ and\ \citenamefont {Nandkishore}}]{frc}%
  \BibitemOpen
  \bibfield  {author} {\bibinfo {author} {\bibfnamefont {S.}~\bibnamefont
  {Pai}}, \bibinfo {author} {\bibfnamefont {M.}~\bibnamefont {Pretko}}, \ and\
  \bibinfo {author} {\bibfnamefont {R.~M.}\ \bibnamefont {Nandkishore}},\
  }\href {\doibase 10.1103/PhysRevX.9.021003} {\bibfield  {journal} {\bibinfo
  {journal} {Phys. Rev. X}\ }\textbf {\bibinfo {volume} {9}},\ \bibinfo {pages}
  {021003} (\bibinfo {year} {2019})}\BibitemShut {NoStop}%
\bibitem [{\citenamefont {Ma}\ and\ \citenamefont
  {Pretko}(2018)}]{ma2018higher}%
  \BibitemOpen
  \bibfield  {author} {\bibinfo {author} {\bibfnamefont {H.}~\bibnamefont
  {Ma}}\ and\ \bibinfo {author} {\bibfnamefont {M.}~\bibnamefont {Pretko}},\
  }\href {\doibase 10.1103/PhysRevB.98.125105} {\bibfield  {journal} {\bibinfo
  {journal} {Phys. Rev. B}\ }\textbf {\bibinfo {volume} {98}},\ \bibinfo
  {pages} {125105} (\bibinfo {year} {2018})}\BibitemShut {NoStop}%
\bibitem [{\citenamefont
  {Pretko}(2017{\natexlab{b}})}]{pretko2017subdimensional}%
  \BibitemOpen
  \bibfield  {author} {\bibinfo {author} {\bibfnamefont {M.}~\bibnamefont
  {Pretko}},\ }\href@noop {} {\bibfield  {journal} {\bibinfo  {journal}
  {Physical Review B}\ }\textbf {\bibinfo {volume} {95}},\ \bibinfo {pages}
  {115139} (\bibinfo {year} {2017}{\natexlab{b}})}\BibitemShut {NoStop}%
\bibitem [{\citenamefont {Pretko}(2017{\natexlab{c}})}]{pretko2017generalized}%
  \BibitemOpen
  \bibfield  {author} {\bibinfo {author} {\bibfnamefont {M.}~\bibnamefont
  {Pretko}},\ }\href@noop {} {\bibfield  {journal} {\bibinfo  {journal}
  {Physical Review B}\ }\textbf {\bibinfo {volume} {96}},\ \bibinfo {pages}
  {035119} (\bibinfo {year} {2017}{\natexlab{c}})}\BibitemShut {NoStop}%
\bibitem [{\citenamefont {Ma}\ \emph {et~al.}(2018{\natexlab{b}})\citenamefont
  {Ma}, \citenamefont {Hermele},\ and\ \citenamefont {Chen}}]{ma2018fracton}%
  \BibitemOpen
  \bibfield  {author} {\bibinfo {author} {\bibfnamefont {H.}~\bibnamefont
  {Ma}}, \bibinfo {author} {\bibfnamefont {M.}~\bibnamefont {Hermele}}, \ and\
  \bibinfo {author} {\bibfnamefont {X.}~\bibnamefont {Chen}},\ }\href@noop {}
  {\bibfield  {journal} {\bibinfo  {journal} {Phys. Rev. B}\ }\textbf {\bibinfo
  {volume} {98}},\ \bibinfo {pages} {035111} (\bibinfo {year}
  {2018}{\natexlab{b}})}\BibitemShut {NoStop}%
\bibitem [{\citenamefont {Bulmash}\ and\ \citenamefont
  {Barkeshli}(2018{\natexlab{a}})}]{bulmash2018higgs}%
  \BibitemOpen
  \bibfield  {author} {\bibinfo {author} {\bibfnamefont {D.}~\bibnamefont
  {Bulmash}}\ and\ \bibinfo {author} {\bibfnamefont {M.}~\bibnamefont
  {Barkeshli}},\ }\href@noop {} {\bibfield  {journal} {\bibinfo  {journal}
  {Phys. Rev. B}\ }\textbf {\bibinfo {volume} {97}},\ \bibinfo {pages} {235112}
  (\bibinfo {year} {2018}{\natexlab{a}})}\BibitemShut {NoStop}%
\bibitem [{\citenamefont {Bulmash}\ and\ \citenamefont
  {Barkeshli}(2018{\natexlab{b}})}]{bulmash2018generalized}%
  \BibitemOpen
  \bibfield  {author} {\bibinfo {author} {\bibfnamefont {D.}~\bibnamefont
  {Bulmash}}\ and\ \bibinfo {author} {\bibfnamefont {M.}~\bibnamefont
  {Barkeshli}},\ }\href@noop {} {\bibfield  {journal} {\bibinfo  {journal}
  {arXiv preprint arXiv:1806.01855}\ } (\bibinfo {year}
  {2018}{\natexlab{b}})}\BibitemShut {NoStop}%
\bibitem [{\citenamefont {Pretko}(2017{\natexlab{d}})}]{Pretko2017-nt}%
  \BibitemOpen
  \bibfield  {author} {\bibinfo {author} {\bibfnamefont {M.}~\bibnamefont
  {Pretko}},\ }\href@noop {} {\bibfield  {journal} {\bibinfo  {journal} {Phys.
  Rev. B Condens. Matter}\ }\textbf {\bibinfo {volume} {96}},\ \bibinfo {pages}
  {125151} (\bibinfo {year} {2017}{\natexlab{d}})}\BibitemShut {NoStop}%
\bibitem [{\citenamefont {{You}}\ \emph {et~al.}(2019)\citenamefont {{You}},
  \citenamefont {{Devakul}}, \citenamefont {{Sondhi}},\ and\ \citenamefont
  {{Burnell}}}]{fractoncs}%
  \BibitemOpen
  \bibfield  {author} {\bibinfo {author} {\bibfnamefont {Y.}~\bibnamefont
  {{You}}}, \bibinfo {author} {\bibfnamefont {T.}~\bibnamefont {{Devakul}}},
  \bibinfo {author} {\bibfnamefont {S.~L.}\ \bibnamefont {{Sondhi}}}, \ and\
  \bibinfo {author} {\bibfnamefont {F.~J.}\ \bibnamefont {{Burnell}}},\
  }\href@noop {} {\bibfield  {journal} {\bibinfo  {journal} {arXiv e-prints}\
  ,\ \bibinfo {eid} {arXiv:1904.11530}} (\bibinfo {year} {2019})},\ \Eprint
  {http://arxiv.org/abs/1904.11530} {arXiv:1904.11530 [cond-mat.str-el]}
  \BibitemShut {NoStop}%
\bibitem [{\citenamefont {Gromov}(2018)}]{gromov2018towards}%
  \BibitemOpen
  \bibfield  {author} {\bibinfo {author} {\bibfnamefont {A.}~\bibnamefont
  {Gromov}},\ }\href@noop {} {\bibfield  {journal} {\bibinfo  {journal} {arXiv
  preprint arXiv:1812.05104}\ } (\bibinfo {year} {2018})}\BibitemShut {NoStop}%
\bibitem [{\citenamefont {Xu}\ and\ \citenamefont {Fisher}(2007)}]{xu2007bond}%
  \BibitemOpen
  \bibfield  {author} {\bibinfo {author} {\bibfnamefont {C.}~\bibnamefont
  {Xu}}\ and\ \bibinfo {author} {\bibfnamefont {M.~P.}\ \bibnamefont
  {Fisher}},\ }\href@noop {} {\bibfield  {journal} {\bibinfo  {journal}
  {Physical Review B}\ }\textbf {\bibinfo {volume} {75}},\ \bibinfo {pages}
  {104428} (\bibinfo {year} {2007})}\BibitemShut {NoStop}%
\bibitem [{\citenamefont {Paramekanti}\ \emph {et~al.}(2002)\citenamefont
  {Paramekanti}, \citenamefont {Balents},\ and\ \citenamefont
  {Fisher}}]{paramekanti2002ring}%
  \BibitemOpen
  \bibfield  {author} {\bibinfo {author} {\bibfnamefont {A.}~\bibnamefont
  {Paramekanti}}, \bibinfo {author} {\bibfnamefont {L.}~\bibnamefont
  {Balents}}, \ and\ \bibinfo {author} {\bibfnamefont {M.~P.}\ \bibnamefont
  {Fisher}},\ }\href@noop {} {\bibfield  {journal} {\bibinfo  {journal}
  {Physical Review B}\ }\textbf {\bibinfo {volume} {66}},\ \bibinfo {pages}
  {054526} (\bibinfo {year} {2002})}\BibitemShut {NoStop}%
\bibitem [{\citenamefont {Tay}\ and\ \citenamefont
  {Motrunich}(2010)}]{tay2010possible}%
  \BibitemOpen
  \bibfield  {author} {\bibinfo {author} {\bibfnamefont {T.}~\bibnamefont
  {Tay}}\ and\ \bibinfo {author} {\bibfnamefont {O.~I.}\ \bibnamefont
  {Motrunich}},\ }\href@noop {} {\bibfield  {journal} {\bibinfo  {journal}
  {Physical review letters}\ }\textbf {\bibinfo {volume} {105}},\ \bibinfo
  {pages} {187202} (\bibinfo {year} {2010})}\BibitemShut {NoStop}%
\bibitem [{\citenamefont {Tay}\ \emph {et~al.}(2011)\citenamefont {Tay},
  \citenamefont {Motrunich} \emph {et~al.}}]{tay2011possible}%
  \BibitemOpen
  \bibfield  {author} {\bibinfo {author} {\bibfnamefont {T.}~\bibnamefont
  {Tay}}, \bibinfo {author} {\bibfnamefont {O.~I.}\ \bibnamefont {Motrunich}},
  \emph {et~al.},\ }\href@noop {} {\bibfield  {journal} {\bibinfo  {journal}
  {Physical Review B}\ }\textbf {\bibinfo {volume} {83}},\ \bibinfo {pages}
  {205107} (\bibinfo {year} {2011})}\BibitemShut {NoStop}%
\bibitem [{\citenamefont {Mishmash}\ \emph {et~al.}(2011)\citenamefont
  {Mishmash}, \citenamefont {Block}, \citenamefont {Kaul}, \citenamefont
  {Sheng}, \citenamefont {Motrunich},\ and\ \citenamefont
  {Fisher}}]{mishmash2011bose}%
  \BibitemOpen
  \bibfield  {author} {\bibinfo {author} {\bibfnamefont {R.~V.}\ \bibnamefont
  {Mishmash}}, \bibinfo {author} {\bibfnamefont {M.~S.}\ \bibnamefont {Block}},
  \bibinfo {author} {\bibfnamefont {R.~K.}\ \bibnamefont {Kaul}}, \bibinfo
  {author} {\bibfnamefont {D.}~\bibnamefont {Sheng}}, \bibinfo {author}
  {\bibfnamefont {O.~I.}\ \bibnamefont {Motrunich}}, \ and\ \bibinfo {author}
  {\bibfnamefont {M.~P.}\ \bibnamefont {Fisher}},\ }\href@noop {} {\bibfield
  {journal} {\bibinfo  {journal} {Physical Review B}\ }\textbf {\bibinfo
  {volume} {84}},\ \bibinfo {pages} {245127} (\bibinfo {year}
  {2011})}\BibitemShut {NoStop}%
\bibitem [{\citenamefont {Lai}\ \emph {et~al.}(2013{\natexlab{a}})\citenamefont
  {Lai}, \citenamefont {Yang},\ and\ \citenamefont
  {Bonesteel}}]{lai2013violation}%
  \BibitemOpen
  \bibfield  {author} {\bibinfo {author} {\bibfnamefont {H.-H.}\ \bibnamefont
  {Lai}}, \bibinfo {author} {\bibfnamefont {K.}~\bibnamefont {Yang}}, \ and\
  \bibinfo {author} {\bibfnamefont {N.}~\bibnamefont {Bonesteel}},\ }\href@noop
  {} {\bibfield  {journal} {\bibinfo  {journal} {Physical review letters}\
  }\textbf {\bibinfo {volume} {111}},\ \bibinfo {pages} {210402} (\bibinfo
  {year} {2013}{\natexlab{a}})}\BibitemShut {NoStop}%
\bibitem [{\citenamefont {Lieb}\ \emph {et~al.}(1961)\citenamefont {Lieb},
  \citenamefont {Schultz},\ and\ \citenamefont {Mattis}}]{lsm}%
  \BibitemOpen
  \bibfield  {author} {\bibinfo {author} {\bibfnamefont {E.}~\bibnamefont
  {Lieb}}, \bibinfo {author} {\bibfnamefont {T.}~\bibnamefont {Schultz}}, \
  and\ \bibinfo {author} {\bibfnamefont {D.}~\bibnamefont {Mattis}},\ }\href
  {\doibase https://doi.org/10.1016/0003-4916(61)90115-4} {\bibfield  {journal}
  {\bibinfo  {journal} {Annals of Physics}\ }\textbf {\bibinfo {volume} {16}},\
  \bibinfo {pages} {407 } (\bibinfo {year} {1961})}\BibitemShut {NoStop}%
\bibitem [{\citenamefont {Hastings}(2004)}]{hastings}%
  \BibitemOpen
  \bibfield  {author} {\bibinfo {author} {\bibfnamefont {M.~B.}\ \bibnamefont
  {Hastings}},\ }\href {\doibase 10.1103/PhysRevB.69.104431} {\bibfield
  {journal} {\bibinfo  {journal} {Phys. Rev. B}\ }\textbf {\bibinfo {volume}
  {69}},\ \bibinfo {pages} {104431} (\bibinfo {year} {2004})}\BibitemShut
  {NoStop}%
\bibitem [{\citenamefont {Oshikawa}(2000)}]{oshikawa}%
  \BibitemOpen
  \bibfield  {author} {\bibinfo {author} {\bibfnamefont {M.}~\bibnamefont
  {Oshikawa}},\ }\href {\doibase 10.1103/PhysRevLett.84.1535} {\bibfield
  {journal} {\bibinfo  {journal} {Phys. Rev. Lett.}\ }\textbf {\bibinfo
  {volume} {84}},\ \bibinfo {pages} {1535} (\bibinfo {year}
  {2000})}\BibitemShut {NoStop}%
\bibitem [{\citenamefont {Pankov}\ \emph
  {et~al.}(2007{\natexlab{b}})\citenamefont {Pankov}, \citenamefont
  {Moessner},\ and\ \citenamefont {Sondhi}}]{pankov2007resonating}%
  \BibitemOpen
  \bibfield  {author} {\bibinfo {author} {\bibfnamefont {S.}~\bibnamefont
  {Pankov}}, \bibinfo {author} {\bibfnamefont {R.}~\bibnamefont {Moessner}}, \
  and\ \bibinfo {author} {\bibfnamefont {S.~L.}\ \bibnamefont {Sondhi}},\
  }\href@noop {} {\bibfield  {journal} {\bibinfo  {journal} {Physical Review
  B}\ }\textbf {\bibinfo {volume} {76}},\ \bibinfo {pages} {104436} (\bibinfo
  {year} {2007}{\natexlab{b}})}\BibitemShut {NoStop}%
\bibitem [{\citenamefont {Fradkin}\ \emph {et~al.}(2004)\citenamefont
  {Fradkin}, \citenamefont {Huse}, \citenamefont {Moessner}, \citenamefont
  {Oganesyan},\ and\ \citenamefont {Sondhi}}]{cantor}%
  \BibitemOpen
  \bibfield  {author} {\bibinfo {author} {\bibfnamefont {E.}~\bibnamefont
  {Fradkin}}, \bibinfo {author} {\bibfnamefont {D.~A.}\ \bibnamefont {Huse}},
  \bibinfo {author} {\bibfnamefont {R.}~\bibnamefont {Moessner}}, \bibinfo
  {author} {\bibfnamefont {V.}~\bibnamefont {Oganesyan}}, \ and\ \bibinfo
  {author} {\bibfnamefont {S.~L.}\ \bibnamefont {Sondhi}},\ }\href {\doibase
  10.1103/PhysRevB.69.224415} {\bibfield  {journal} {\bibinfo  {journal} {Phys.
  Rev. B}\ }\textbf {\bibinfo {volume} {69}},\ \bibinfo {pages} {224415}
  (\bibinfo {year} {2004})}\BibitemShut {NoStop}%
\bibitem [{Note1()}]{Note1}%
  \BibitemOpen
  \bibinfo {note} {$\protect \mathaccentV {hat}05E{n}_-=s_z+1/2$ is the
  conjugate variable of $\theta _-$.}\BibitemShut {Stop}%
\bibitem [{\citenamefont {Kohno}\ \emph {et~al.}(2007)\citenamefont {Kohno},
  \citenamefont {Starykh},\ and\ \citenamefont {Balents}}]{kohno2007spinons}%
  \BibitemOpen
  \bibfield  {author} {\bibinfo {author} {\bibfnamefont {M.}~\bibnamefont
  {Kohno}}, \bibinfo {author} {\bibfnamefont {O.~A.}\ \bibnamefont {Starykh}},
  \ and\ \bibinfo {author} {\bibfnamefont {L.}~\bibnamefont {Balents}},\
  }\href@noop {} {\bibfield  {journal} {\bibinfo  {journal} {Nature Physics}\
  }\textbf {\bibinfo {volume} {3}},\ \bibinfo {pages} {790} (\bibinfo {year}
  {2007})}\BibitemShut {NoStop}%
\bibitem [{\citenamefont {Lai}\ \emph {et~al.}(2013{\natexlab{b}})\citenamefont
  {Lai}, \citenamefont {Yang},\ and\ \citenamefont
  {Bonesteel}}]{boseentanglement}%
  \BibitemOpen
  \bibfield  {author} {\bibinfo {author} {\bibfnamefont {H.-H.}\ \bibnamefont
  {Lai}}, \bibinfo {author} {\bibfnamefont {K.}~\bibnamefont {Yang}}, \ and\
  \bibinfo {author} {\bibfnamefont {N.~E.}\ \bibnamefont {Bonesteel}},\ }\href
  {\doibase 10.1103/PhysRevLett.111.210402} {\bibfield  {journal} {\bibinfo
  {journal} {Phys. Rev. Lett.}\ }\textbf {\bibinfo {volume} {111}},\ \bibinfo
  {pages} {210402} (\bibinfo {year} {2013}{\natexlab{b}})}\BibitemShut
  {NoStop}%
\bibitem [{\citenamefont {Wolf}(2006)}]{wolf}%
  \BibitemOpen
  \bibfield  {author} {\bibinfo {author} {\bibfnamefont {M.~M.}\ \bibnamefont
  {Wolf}},\ }\href {\doibase 10.1103/PhysRevLett.96.010404} {\bibfield
  {journal} {\bibinfo  {journal} {Phys. Rev. Lett.}\ }\textbf {\bibinfo
  {volume} {96}},\ \bibinfo {pages} {010404} (\bibinfo {year}
  {2006})}\BibitemShut {NoStop}%
\bibitem [{\citenamefont {Gioev}\ and\ \citenamefont {Klich}(2006)}]{widom}%
  \BibitemOpen
  \bibfield  {author} {\bibinfo {author} {\bibfnamefont {D.}~\bibnamefont
  {Gioev}}\ and\ \bibinfo {author} {\bibfnamefont {I.}~\bibnamefont {Klich}},\
  }\href {\doibase 10.1103/PhysRevLett.96.100503} {\bibfield  {journal}
  {\bibinfo  {journal} {Phys. Rev. Lett.}\ }\textbf {\bibinfo {volume} {96}},\
  \bibinfo {pages} {100503} (\bibinfo {year} {2006})}\BibitemShut {NoStop}%
\bibitem [{\citenamefont {Swingle}(2010)}]{swingle}%
  \BibitemOpen
  \bibfield  {author} {\bibinfo {author} {\bibfnamefont {B.}~\bibnamefont
  {Swingle}},\ }\href {\doibase 10.1103/PhysRevLett.105.050502} {\bibfield
  {journal} {\bibinfo  {journal} {Phys. Rev. Lett.}\ }\textbf {\bibinfo
  {volume} {105}},\ \bibinfo {pages} {050502} (\bibinfo {year}
  {2010})}\BibitemShut {NoStop}%
\bibitem [{Note2()}]{Note2}%
  \BibitemOpen
  \bibinfo {note} {The cube $\protect \bm {r}$ is the adjacent cube of site
  $\protect \bm {r}$ along the $(111)$ direction.}\BibitemShut {Stop}%
\end{thebibliography}
\end{document}